\documentclass[]{ws-mplb}
\usepackage{amsfonts,amssymb,amsmath}
\usepackage{graphicx}
\usepackage[super,sort,compress]{cite} 
\usepackage{url}

\newcommand{\bsigma}{\boldsymbol{\sigma}}
\newcommand{\bdelta}{\boldsymbol{\delta}}
\newcommand{\br}{\mathbf{r}}
\newcommand{\bk}{\mathbf{k}}
\newcommand{\bA}{\mathbf{A}}
\newcommand{\bB}{\mathbf{B}}

\begin{document}

\markboth{Dawei Zhai and Nancy Sandler}{Electron dynamics in strained graphene}

%
\catchline{}{}{}{}{}
%

\title{Electron dynamics in strained graphene}

\author{Dawei Zhai\footnote{Present address:  Department of Physics, The University of Hong Kong, Hong Kong, China}}

\address{Department of Physics and Astronomy, Ohio University\\
Athens Ohio, 45701, USA\\
dzhai@hku.hk}

\author{Nancy Sandler}

\address{Department of Physics and Astronomy, Ohio University\\
Athens Ohio, 45701, USA\\
sandler@ohio.edu}

\maketitle

\begin{history}
\received{(Day Month Year)}
\revised{(Day Month Year)}
\end{history}

\begin{abstract}
The paper presents a theoretical description of the effects of strain induced by out-of-plane deformations on charge distributions and transport on graphene. A review of a continuum model for electrons using the Dirac formalism is complemented with elasticity theory to represent strain fields. The resulting model is cast in terms of scalar and pseudo-magnetic fields that control electron dynamics.
Two distinct geometries, a bubble and a fold, are chosen to represent the most commonly observed deformations in experimental settings. It is shown that local charge accumulation regions appear in deformed areas, with a peculiar charge distribution that favors occupation of one sublattice only. This unique phenomenon that allows to distinguish each carbon atom in the unit cell, is the manifestation of a sublattice symmetry broken phase. For specific parameters, resonant states appear in localized charged regions, as shown by the emergence of discrete levels in band structure calculations. These findings are presented in terms of intuitive pictures that exploit analogies with confinement produced by square barriers. In addition, electron currents through strained regions are spatially separated into their valley components, making possible the manipulation of electrons with different valley indices. The degree of valley filtering (or polarization) for a specific system can be controlled by properly designing the strained area. The comparison between efficiencies of filters built with this type of geometries identifies extended deformations as better valley filters. A proposal for their experimental implementations as component of devices, and a discussion for potential observation of novel physics in strained structures are presented at the end of the article.
\end{abstract}

\section{Introduction}
With increasing miniaturization and stringent requirements for reduced heat dissipation, the development of new technologies propels the continuous search for more perfect materials (free from impurities and defects) and novel mechanisms to control electronic currents with greater precision. The isolation of graphene and other two-dimensional compounds in 2004\cite{KostyaScience2004} gave access not only to low-dimensional versions of well-known materials but also to single-layer manipulation methods, together with a prodigious number of potential new compounds obtained via multiple single-layer combinations\cite{GeimLegos,KostyaLegos}. In particular, advances on sample fabrication have allowed the observation of new phenomena\cite{ValleyHallMak,GrapheneHofstadter,HuntGrapheneHofstadter,SuperconductivityTwistedBilayer,CaoYuanCorrelated,VeselagoLensing} and contributed to a better understanding of the intrinsic electronic properties of mono-layer systems. 

Of all these new materials, graphene, has become a test-bench for studies of electronic and transport properties, making it the best understood material from the group\cite{CastroNeto2009,GrapheneTransportRMP,Confinement_A_barrier_RMP2010,GoerbigRMP,ValleyIsotropicNotation,GaugeFieldPhysRep,Strained2DPhysRep2016}. Nowhitstanding the vast amount of work carried on, a wide range of new findings have been revealed by the improvement of experimental methods used to manipulate the material with greater precision. This is particularly evident for work involving deformed samples, in which electronic properties are fundamentally altered due to the presence of strain.\cite{GaugeFieldPhysRep,Strained2DPhysRep2016} The realization that strain exists in supported and suspended samples, and can be introduced by substrates as well as by imaging techniques,\cite{Substrate1,Substrate2,Substrate3,Substrate4,YuhangNanoLett2017,Thermal1,Thermal2,Thermal3,Moire1,Moire2,MarkusNanoLett2017,STM_tip_bubble_Stroscio,STM_tip_bubble_Stroscio_PRB,Mechanical2,Mechanical3,Pressure,Pressure2,Pressure3,Pressure4,STM_indentation} had opened the door to a large number of proposals that take advantage of these effects, giving rise to the new field of `straintronics'\cite{StraintronicsReview2018,Strained2DPhysRep2016}.

As strain appears to be ever-present, a clear understanding of its effects becomes essential for appropriate strain-designed engineering. Almost invariably, current fabrication methods and manipulation techniques render samples with strain-induced deformations associated with bent portions of the sample. In these deformed sections, atoms exhibit out-of-plane displacements as the sample distorts to minimize energy constraints. Usually, these frequently observed out-of-plane deformations are produced either locally, e.g., impurities trapped between the graphene membrane and the substrate,\cite{Crommie,MarkusNanoLett2017} or in more extended sections of the sample, as is the case of folds and wrinkles.\cite{YuhangNanoLett2017,FoldNatCommun2015,FoldPRB2015,Substrate1,Substrate2,Substrate3,Thermal1,Thermal2,Mechanical3} 

The purpose of this article is to present a comprehensive account of the latest developments, and discuss experimental realizations that take advantage of existing strain in samples and may lead to potential device applications. As described below, strain modifies charge distributions, and consequently electronic transport, in very specific ways making it an exciting new tool for tailored material properties.

The manuscript is based on Ref.~\citen{ZhaiDissertation} and organized as follows. Section~\ref{Sect:Expt_strain_engineering} reviews the experimental methods used to induce and control deformations on graphene membranes. These works provide evidence for the statement above regarding the ubiquity of out-of plane deformations. In Sect.~\ref{Sect:TheoryBackground} we present the basic elements of the model for a continuum description of electron dynamics in graphene based on the Dirac equation. We introduce the concept of scalar and pseudo-magnetic fields, that generate an intuitive description of changes in charge distribution and transport in the presence of deformations, within an elegant and concise formalism. The section closes with the specific formulation of the two model geometries described above.
Section~\ref{Sect:LDOS} presents an analysis of the effects these deformations have on the charge distribution in graphene. The mechanism by which deformed regions confine charges, a fact that is already confirmed by experimental results, is described in detail. Surprisingly, the confinement also occurs in a very peculiar fashion: the inhomogeneous nature of the deformation produces an uneven charge distribution at the level of the unit cell, i.e., electric charge is re-arranged between the two identical carbon atoms following a precise pattern that maintains the balance between sublattices at the global level. 
This phenomena, dubbed sublattice symmetry breaking, has been observed in several experimental settings and is used nowadays as a tool to quantify the amount of strain present in a given sample.
In Sect.~\ref{Sect:ValleyFilter}, the role of deformations on charge transport are discussed. The section introduces the functioning principles of valley filters and valley polarizers, including advantages and disadvantages for their implementation and use in applications. As done in previous works, we characterize the efficiency of each valley filter taking into account the amount of valley separation introduced and also the magnitude of the resulting polarized current. 
The review closes with Sect.~\ref{Sect:Future} that contains experimental proposals and future directions in strained graphene.

\section{Experimental methods of strain engineering in graphene}\label{Sect:Expt_strain_engineering}

\begin{figure}[th]
	\centerline{\includegraphics[width=5in]{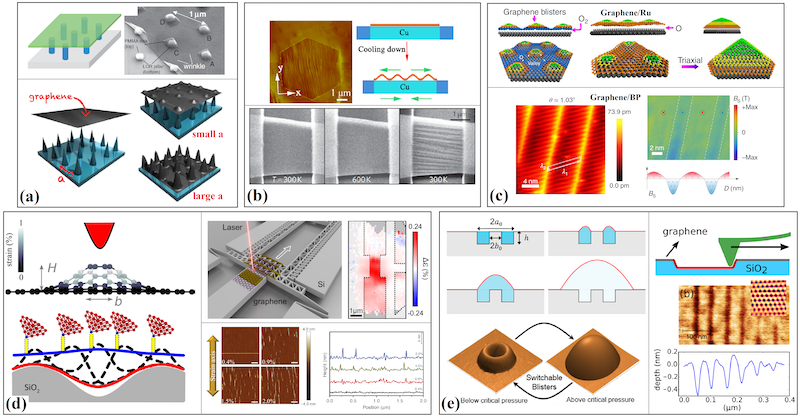}}
	\vspace*{8pt}
	\caption{Practical approaches for strain engineering in graphene (Gray lines in each panel separate figures adapted from different references). (a-c) Substrate-based approaches. Panel (a) presents substrate engineering with nanopillars fabricated from dielectric materials, on top of Si substrates. Graphene nanobubbles form directly on top of each pillar, while wrinkles form between pillars.\cite{Substrate1} The geometry of the deformation can be tuned via changes in shape, size and distance (density) of the individual substrate units.\cite{Substrate2} Panel (b) shows thermal control of deformations by annealing. Wrinkles form when cooling the system due to opposite TEC values of graphene and substrate.\cite{Thermal1} The geometry of the wrinkles can be tuned via temperature, shape (e.g. trench vs circular hole), and size of the substrate.\cite{Thermal2} Panel (c) illustrates how lattice mismatch between graphene and the substrate can be exploited to generate strain. The upper row shows graphene deposited on ruthenium. Small blisters can be observed in the picture that are formed after deposition. These blisters can then be turned into larger deformations by merging through e.g. oxygen injection.\cite{Moire1} The lower row shows a graphene-black phosphorus superlattice and its simulated pseudo-magnetic field (see text).\cite{Moire2} (d-e) Mechanical approaches. Panel (d) shows the generation of nanobubbles using an STM tip (left column)\cite{MarkusNanoLett2017}. In this geometry the bubble is created by electrostatic attraction between graphene and the tip and the deformation  moves as the tip scans the sample (dashed curves). A more controlled strain engineering via anisotropic straining of the sample is shown in the right column\cite{Mechanical2,Mechanical3}. The top right figure shows the corresponding strain distribution while the bottom right figure shows height profiles of out-of-plane deformations. Panel (e) presents strain engineering by combining substrate fabrication and external forces. Figures in the left column show how to control deformation by tuning the pressure\cite{Pressure}, while those in the right column are examples of measurements done on suspended graphene after applying a force with an STM tip\cite{STM_indentation}. \label{Fig:Strain_engineering_expt}}
\end{figure}

Thermodynamically speaking, perfectly flat 2D materials are unstable. Therefore, deformations (e.g. wrinkles), naturally found in graphene samples, help to stabilize the system.\cite{GrapheneDeformationReviewMaterToday} From the point of view of applications, it is desirable to control the conditions under which deformations are created as well as their characteristic parameters. Several approaches have been developed to manipulate strain via deformations in graphene, and we discuss a few representatives ones in this section.\cite{Substrate1,Substrate2,Substrate3,Substrate4,YuhangNanoLett2017,Thermal1,Thermal2,Thermal3,Moire1,Moire2,MarkusNanoLett2017,STM_tip_bubble_Stroscio,STM_tip_bubble_Stroscio_PRB,Mechanical2,Mechanical3,Pressure,Pressure2,Pressure3,Pressure4,STM_indentation}

In the following we classify the strain/deformation engineering methods into two groups: (i) substrate-based approaches, where the substrates determine the deformation/strain profile; and (ii) mechanical approaches, where direct manipulation of the sample using external forces, determines the resulting strain profile.

(i) \textit{Substrate-based approaches}.
The most common method, known as substrate engineering, consists of the fabrication of structures with definite shapes (e.g. cylindrical, spherical, pyramidal, etc) on top of which graphene is deposited (Fig.~\ref{Fig:Strain_engineering_expt}(a) upper row).\cite{Substrate1,Substrate2,Substrate3,Substrate4,YuhangNanoLett2017}. Graphene is deformed in the area that lies directly on top of these structures and in the areas between adjacent units (e.g. wrinkles formed between nanopillars). By controlling the geometry, size and arrangement (e.g. density) of individual units, a precise control of strain profiles can be achieved (Fig.~\ref{Fig:Strain_engineering_expt}(a) lower row).

Another method is based on the differences between thermal expansion coefficients (TEC) between graphene and substrate (Fig.~\ref{Fig:Strain_engineering_expt}(b)).\cite{Thermal1,Thermal2,Thermal3} Graphene has a negative TEC, while most substrates have positive values. When graphene is grown or deposited on top of a substrate, cooling processes produce the expansion of graphene and the compression of the substrate. These opposite trends, in combination with the membrane's relaxation, naturally generate out-of-plane deformations that reduce the total elastic energy of the system. (Fig.~\ref{Fig:Strain_engineering_expt}(b) upper row).\cite{Thermal1,Thermal3} Similarly, it is possible to induce ripples in a sample of graphene suspended over a trench substrate by annealing. In this situation, the formed ripples align in the direction perpendicular to the trench (Fig.~\ref{Fig:Strain_engineering_expt}(b) lower row).\cite{Thermal2} In this setup, the geometry of the ripples is tunable through the temperature and shape of the underlying substrate (e.g. trench, hole, etc).

One can also employ the lattice mismatch between graphene and the substrate to induce strain (Fig.~\ref{Fig:Strain_engineering_expt}(c)).\cite{Moire1,Moire2} For instance, small blisters form due to the Moir\'{e} pattern appearing when graphene is deposited on Ruthenium or Germanium. Larger deformations with various shapes can be achieved through selectively merging several blisters with e.g. Oxygen injection (Fig.~\ref{Fig:Strain_engineering_expt}(c) upper row). Furthermore, one can engineer the strain profile by selecting materials with different lattice constants and/or controlling the twist angle between graphene and the substrate (Fig.~\ref{Fig:Strain_engineering_expt}(c) lower row).

(ii) \textit{Mechanical approaches}.
Obviously the most straightforward method under this category is to lift the membrane with a `tweezer', e.g. scanning tunneling microscope (STM) tip (Fig.~\ref{Fig:Strain_engineering_expt}(d) left column).\cite{MarkusNanoLett2017,STM_tip_bubble_Stroscio,STM_tip_bubble_Stroscio_PRB} One can even move the deformation dynamically to different locations. However, it is challenging to obtain sizable deformations, or control their shape. Furthermore, any measurement (e.g. transport) is influenced by the presence of the tip. A more promising approach involves depositing graphene on a flexible substrate and stretch/compress the system with external forces, which can not only yield sizable and tunable in-plane strain, but also out-of-plane deformations (Fig.~\ref{Fig:Strain_engineering_expt}(d) right column).\cite{Mechanical2,Mechanical3}

Another mechanical approach involves applying forces on suspended graphene (Fig.~\ref{Fig:Strain_engineering_expt}(e)).\cite{Pressure,Pressure2,Pressure3,Pressure4,STM_indentation} In these settings, graphene is deposited on a substrate with indentations (holes, rings, trenches, etc), that are filled with gas. As the gas pressure is increased, graphene is deformed (Fig.~\ref{Fig:Strain_engineering_expt}(e) left column). This setup also offers a controlled approach for the application of forces via STM tips (Fig.~\ref{Fig:Strain_engineering_expt}(e) right column). In both cases, the profile of the deformation can be controlled via size and shape of the indentation as well as the magnitude of the external force (e.g. by changing the pressure).

As shown in Fig.~\ref{Fig:Strain_engineering_expt}, deformations in graphene usually exhibit local bubble-like structure or extended fold/wrinkle-like geometry. Consequently, it is instructive to focus on these two typical geometries, represented by Gaussian bubbles and folds throughout this paper.

\section{Theory background}\label{Sect:TheoryBackground}
In this section we present the Dirac formalism used to describe the electronic properties of strained graphene. This continuum approach incorporates the effects of strain by introducing a pseudo-vector and a scalar potentials. As will be shown later, such an effective model offers intuitive pictures for understanding the electron dynamics in strained graphene.

Arguably the most attractive aspect of graphene is the low energy relativistic behavior of electrons in the two inequivalent valleys labeled $K$ and $K'$. Because a large momentum transfer is necessary to mix them, this valley index can be used as a quantum number in the presence of weak disorder. As a consequence, for reasonably pristine samples, electron dynamics in each valley is well described by a massless Dirac equation:
\begin{equation}
	H_{K,K'}=v_F\bsigma\cdot\textbf{p}
\end{equation}
where we have used the valley isotropic basis $(KA,KB,-K'B,K'A)$ representation, with $v_F\approx10^6$ m/s the Fermi velocity~\cite{ValleyIsotropicNotation,CastroNeto2009,GoerbigRMP,GrapheneTransportRMP} and $\bsigma=(\sigma_x,\sigma_y)$ the pseudo-spin. The latter is introduced to reflect the honeycomb lattice structure, which is composed of two inequivalent triangular sublattices usually labeled as $A$ and $B$.  

\begin{figure}[th]
	\centerline{\includegraphics[width=3.5in]{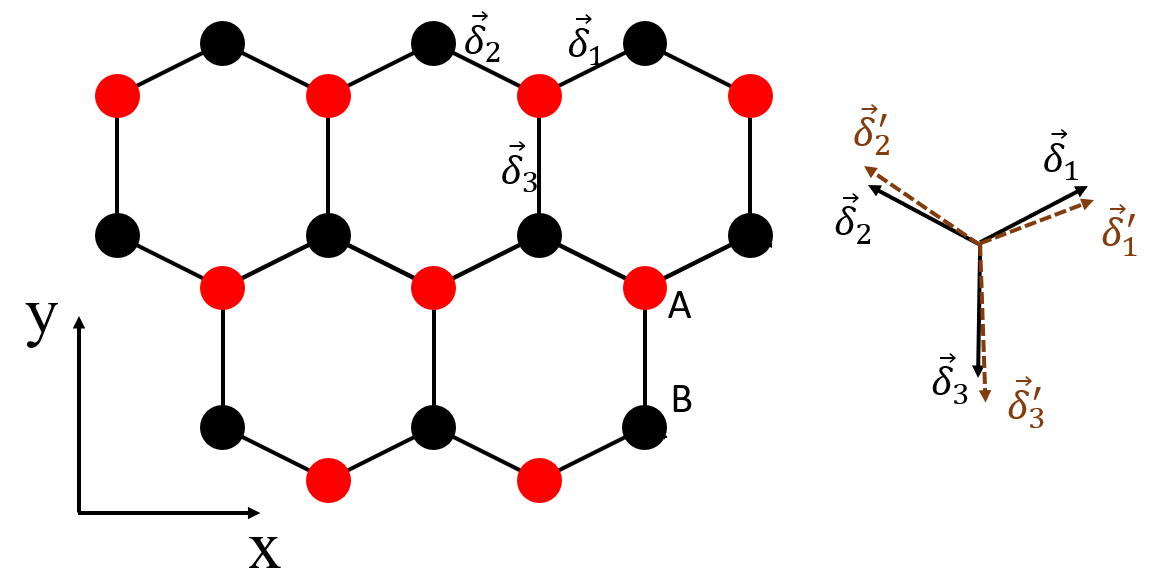}}
	\vspace*{8pt}
	\caption{Schematics of pristine graphene honeycomb lattice and distortion of the nearest-neighbor vectors under strain (from solid to dashed). In our discussions, the $x$ ($y$) direction is assumed to align along the zigzag (armchair) crystalline direction. Figure reproduced from Ref.~\citen{ZhaiDissertation}.\label{Fig:Distorted_lattice}}
\end{figure}

\subsection{Pseudo-vector potential and pseudo-magnetic field in strained graphene}
In the presence of strain, the carbon-carbon distance in each unit cell is modified. If the strain intensity is small, the correction to the nearest-neighbor separation (Fig.~\ref{Fig:Distorted_lattice}) can be approximated as
\begin{equation}
|\bdelta_n'|\approx a+\frac{1}{a}\bdelta_n\cdot\epsilon\cdot\bdelta_n,
\end{equation}
where $a\approx1.4$ \AA \ is the unperturbed carbon-carbon distance, $\bdelta_n$ is the undistorted nearest-neighbor vector, and $\epsilon$ is the strain tensor.\cite{Kitt2012,Kitt2012E}
This effect on electron dynamics is represented by an appropriate change in the hopping energy $t_n$ described by an exponential decay
\begin{equation}
	t_n=te^{-\beta\left(\frac{|\bdelta_n'|}{a}-1\right)},\label{Eq:t_modified}
\end{equation}
where $t\approx 2.7$ eV is the unperturbed hopping energy and $\beta\approx3$~\cite{Ando,GaugeFieldPhysRep,VictorPRB2009,gsGuinea2010,SalvadorPRB2013,SSBMartin,AlexCroy}.

The modified hopping energy in the lattice is represented in the continuum description by a pseudo-vector field $\bA$:
\begin{equation}
H_{\tau}=v_F\bsigma\cdot(\textbf{p}+\tau e\textbf{A}),\label{Eq:H_minimal_substitution}
\end{equation}
where $\tau=\pm$ labels the $K$ and $K'$ valleys respectively, $e$ is the magnitude of the electron charge, and $\bA$ is defined as: 
\begin{equation}
\bA=(A_x,A_y)=-\frac{\hbar\beta}{2ae}(\epsilon_{xx}-\epsilon_{yy},-2\epsilon_{xy}).\label{Eq:A_in_terms_of_epsilon}
\end{equation}
where $\epsilon_{ij}$ is the component of the strain tensor. For a given deformation geometry, one can use the expression $\epsilon_{ij}=\frac{1}{2}(\partial_j u_i+\partial_i u_j+\partial_i h \partial_j h)$ where  $u_{i,j}$ and $h$ are the in-plane and out-of-plane displacement fields~\cite{LandauElasticityBook}.

Eq.~(\ref{Eq:H_minimal_substitution}) is reminiscent of the minimal substitution $(\textbf{p}\rightarrow\textbf{p}-q\textbf{A})$ used to describe the effect of an electromagnetic field (given in terms of its vector potential $\bA$) on  a particle with charge $q$~\cite{RelativisticBookGross}. Therefore, the effect of strain on electron dynamics in graphene can be viewed as similar to that of a magnetic field, which enters the Dirac equation as a `vector potential' $\bA$. Note, however, that the vector potential and the associated magnetic field $\bB_{\tau}=\nabla\times(\tau\bA)$, exhibit opposite signs in the two valleys. Therefore, this strain-induced magnetic field does not break time-reversal symmetry, in contrast to a real magnetic field. To remark on this difference, the magnetic field due to strain is often known as a pseudo-magnetic field, and $\bA$ as the pseudo-vector potential. The presence of pseudo-magnetic fields in strained graphene was first confirmed by scanning tunneling spectroscopy measurements, where the scaling of dI/dV peaks was observed to be consistent with a Landau level description~\cite{Crommie} in the regime where the strain-induced field was spatially uniform on the scale of magnetic length.

\subsection{Scalar potential due to strain}
Apart from the pseudo-vector potential caused by changes in atomic distances, strain also induces a scalar potential
\begin{equation}\label{Eq:ScalarPotentialGeneralForm}
\Phi(\br)=g_s(\epsilon_{xx}+\epsilon_{yy}),
\end{equation}
which is independent of the valley index.
This scalar potential can be understood as originating from the change of charge density due to the variation of the sample area described by $\epsilon_{xx}+\epsilon_{yy}$ \cite{Ando}. The value of $g_s$, which is of the order of a few electron volts, remains controversial both in magnitude and sign~\cite{AlexCroy,SalvadorPRB2013,SalvadorSolidState2013}. The most commonly used value in the literature is $g_s \approx 3$\,eV \cite{gsGuinea2010,SalvadorPRB2013,SSBMartin,AlexCroy} and it is the one adopted in this paper.

\subsection{Some other effects caused by strain}
Apart from the presence of these two potentials, strain is also associated with some other effects, which we briefly discuss here.
\begin{itemlist}
	\item 
	In the above, the low energy Hamiltonian Eq. (\ref{Eq:H_minimal_substitution}) is obtained from an expansion around the Dirac points of pristine graphene. Strictly speaking, the positions of the two Dirac points are shifted by the presence of strain. However, these corrections have been shown to be negligible for small strain, making it possible to use the interpretation of the valley index as an effective quantum number.\cite{Kitt2012,Kitt2012E,FermiVelocityNaumis,Vozmediano2013,PeetersSolidState2013} Furthermore, this effect does not contribute to the pseudo-magnetic field or the scalar potential, and we neglect this effect in the results discussed below.
	\item 
	Apart from the pseudo-vector potential induced by the modification of the carbon-carbon distance, rehybridization of atomic orbitals due to the curvature of the deformed surface introduces an additional pseudo-vector potential $\bA_{\text{curv}}$ with
	\begin{equation}
	\begin{aligned}
	A_{\text{curv,x}}&=-\epsilon_{\pi\pi}\frac{3a^2}{8}\left[\left(\partial_x^2h\right)^2-\left(\partial_y^2h\right)^2\right] \\
	A_{\text{curv,y}}&=\epsilon_{\pi\pi}\frac{3a^2}{4}\left[\partial_{x,y}^2h\left(\partial_x^2h+\partial_y^2h\right)\right]
	\end{aligned},\label{Eq:A_curvature}
	\end{equation}
	where $\epsilon_{\pi\pi}=2.89$ eV and $h$ characterizes the height of the surface corrugation.~\cite{CurvatureEffect}
	Compared to the pseudo-vector potential in Eq.~(\ref{Eq:A_in_terms_of_epsilon}), this curvature-induced potential has an extra $(a/b)^2$ factor, where $b$ is the characteristic width of a deformation. Thus, it only becomes prominent when the spatial extension of the deformation is comparable to the lattice constant~\cite{RainisFold}.
	\item 
	Strain will also renormalize the Fermi velocity through 
	\begin{equation}
	v_F\rightarrow v_F+\Delta v_F= v_F\left[\text{I}+\left(1-\beta\right)\epsilon \right],\label{Eq:vf_renormalization}
	\end{equation}
	the change $\Delta v_F$ is of the order of the strain intensity~\cite{FermiVelocitydeJuan,FermiVelocityNaumis,FermiVelocityLDOS}. One can replace the Fermi velocity in Eq.~(\ref{Eq:H_minimal_substitution}) with the above renormalized value. However, we neglect $\Delta v_F$ as the corresponding term in the Hamiltonian $\Delta v_F \bsigma \cdot \tau e\bA$ is a  higher order correction in strain intensity and the effect is only pronounced in the absence of the pseudo-vector potential.
\end{itemlist}

\subsection{Two representative models: local Gaussian bump and extended Gaussian fold}
\begin{figure}[ht]
	\centering
	\vspace{0.5cm}
	\includegraphics[width=3.5in]{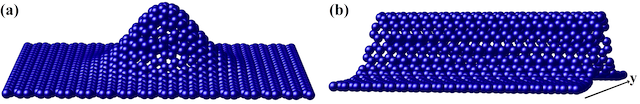}
	\caption[Schematics of the out-of-plane Gaussian (a) bump and (b) fold.]{Schematics of the out-of-plane Gaussian (a) bump and (b) fold. Figure reproduced from Ref.~\citen{MyPaperValleyFilter} with permission.}
	\label{Fig:deformations}
\end{figure}

We have shown that graphene in the presence of strain can be described by
\begin{equation}\label{Eq:Hamilt}
H_{\tau}=v_F\bsigma\cdot\left(\textbf{p}+\tau e \textbf{A}\right) + \Phi(\br),
\end{equation}
where $\bA$ and $\Phi(\br)$ are given in Eqs.~(\ref{Eq:A_in_terms_of_epsilon}) and (\ref{Eq:ScalarPotentialGeneralForm}), respectively. Here we present two models of out-of-plane deformations that are naturally found in experiments and can also be easily engineered (Fig.~\ref{Fig:Strain_engineering_expt}). One is a localized bubble-like shape with circular symmetry, while the other is extended along the $x$ direction resembling a wrinkle or fold (Fig.~\ref{Fig:deformations}).

The two types of deformations are described mathematically by
\begin{equation}
h(\br)=
\begin{cases}
h_0e^{-r^2/b^2}&\text{Gaussian bump}\\
h_0e^{-y^2/b^2}&\text{Gaussian fold}
\end{cases}.
\end{equation}
In these models, in-plane displacements that contribute to the linear terms in the strain tensor are neglected, while the out-of-plane displacements are retained. Neglecting the in-plane displacements is consistent with expected experimental situations where by allowing relaxation, the final atomic positions usually show relatively  smaller in-plane changes. Notice that in-plane displacements with particular spatial dependences can actually contribute to the pseudo-magnetic field, however, such nontrivial in-plane displacement patterns need special and nontrivial sample design/preparation to be produced.

The centrosymmetric Gaussian bump shown in Fig.~\ref{Fig:deformations}(a) produces pseudo-vector and scalar potentials given by
\begin{equation}\label{Eq:bumpfields}
\begin{aligned}
\textbf{A}(\br)&=-\frac{g_v\eta^2}{e v_F}g\left(\frac{r}{b}\right)\left(\cos2\theta,-\sin2\theta\right) \\
\Phi(\br) &= g_s \eta^2 g\left(\frac{r}{b} \right)
\end{aligned},
\end{equation}
where $g_v=\frac{\hbar\beta v_F}{2a}\approx7eV$, $\eta=\frac{h_0}{b}$ is related to the strain strength, $g(z)=2z^2e^{-2z^2}$, and $\theta$ is the polar angle measured with respect to the zigzag crystalline direction.
Analogously, the extended Gaussian fold is translationally invariant along $\hat{x}$ (the zigzag crystalline direction, Fig.~\ref{Fig:deformations}(b)), and produces strain-induced potentials given by
\begin{equation}\label{Eq:foldfields}
\begin{aligned}
\textbf{A}(\br)&=\frac{g_v\eta^2}{ev_F}g\left(\frac{y}{b}\right)\left(1,0\right)\\
\Phi(\br) &= g_s \eta^2 g\left(\frac{y}{b}\right)
\end{aligned}.
\end{equation}

\begin{figure}[ht]
	\centering
	\includegraphics[width=3.5in]{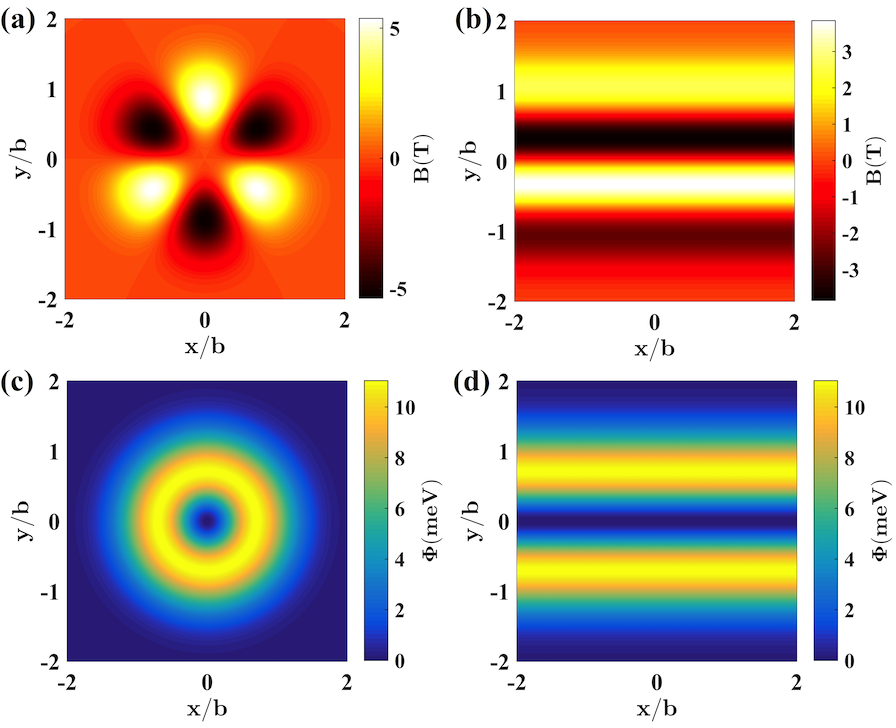}
	\caption{Strain induced pseudo-magnetic field $\textbf{B}_K$ and $\Phi(\br)$ for Gaussian bump and fold. $\textbf{B}_K$ for bump (a), and fold (b). $\Phi(\br)$ for bump (c) and fold (d). Parameters: $\eta=0.1$, $b=15$ nm, $g_s=3$ eV. Figure reproduced from Ref.~\citen{MyPaperValleyFilter} with permission.}
	\label{Fig:B_and_Phi}
\end{figure}

Fig.~\ref{Fig:B_and_Phi} shows typical profiles of the pseudo-magnetic field $\textbf{B}_K=\nabla\times\textbf{A}$ for $K$ valley and scalar potential produced by these two deformations. Results for $\textbf{B}_{K'}$ are obtained by reversing the sign of $\textbf{B}_{K}$. One can see that the pseudo-magnetic field for the bump (Fig.~\ref{Fig:B_and_Phi}(a)) exhibits a flower-like structure with alternating positive and negative regions in the angular direction, while the corresponding scalar potential (Fig.~\ref{Fig:B_and_Phi}(c)) has an isotropic donut-like structure. The fold has translation invariance along its axis. Its pseudo-magnetic field (Fig.~\ref{Fig:B_and_Phi}(b)) exhibits oscillatory behavior with alternating positive and negative regions in the transverse direction, while the scalar potential (Fig.~\ref{Fig:B_and_Phi}(d)) has a double barrier structure.

In the following we focus on these two types of deformations and review interesting phenomena caused by the pseudo-vector and scalar potentials. The goal is to demonstrate that graphene structures with tailored deformations hold potential for various electronic applications.

\section{Effects of strain on charge distribution: sublattice symmetry breaking, confinement, and particle-hole asymmetry}\label{Sect:LDOS}
In this section we analyze the effect of the pseudo-magnetic field and scalar potential on charge distributions. We show that graphene deformations can be employed to manipulate the pseudo-spin degree of freedom, and more importantly, that they can confine charges-- a useful property for building graphene electronic devices.

\subsection{Sublattice symmetry breaking caused by the pseudo-magnetic field}
It is well known that a magnetic field $\bB$ can couple to the spin of an electron through the Zeeman effect, which is represented in the Hamiltonian by the term $-\frac{e\hbar}{2m}\bsigma\cdot\textbf{B}$
in the non-relativistic limit \cite{SaitoGauge,MarkusNanoLett2017}. Note that here $\bsigma$ represents the real spin. Minimization of energy requires the spin to be polarized by the magnetic field, i.e. it aligns with the magnetic field. 

In the case of graphene, $\bsigma$ represents the pseudo-spin, i.e. the sublattice occupancy, and appears in the Hamiltonian coupled to the pseudo-magnetic field through a pseudo-Zeeman term. As is elaborated below, such a coupling has nontrivial consequences as compared to the case of a real magnetic field: the pseudo-spin polarization  is reflected as a sublattice symmetry breaking (SSB) in the local density of states (LDOS). The resulting  effect is a differentiation of the two identical carbon atoms within each unit cell.

We postpone including the strain induced scalar potential for now, and focus on the effects of the pseudo-vector potential first. By squaring $v_F\bsigma\cdot(\textbf{p}+\tau e\textbf{A})\Psi_{\tau}=E\Psi_{\tau}$, one obtains the Pauli equation
\begin{equation}
v_F^2\left[\left(\textbf{p}+\tau e\textbf{A}\right)^2+\tau \hbar e\bsigma\cdot\bB\right]\Psi_{\tau}=E^2\Psi_{\tau},
\end{equation}
where $\tau\hbar e \bsigma\cdot\bB$ acts as a pseudo-Zeeman term~\cite{SSBBlanter,MarkusNanoLett2017,SaitoGauge}. Although the pseudo-Zeeman term as written has opposite signs in the two valleys, its effect on the charge distribution does not cancel between the two valleys. This is clearly seen when recalling basis chosen to represent electron densities: the valley isotropic basis $(KA,KB,-K'B,K'A)$. This choice makes clear that the same pseudo-spin component in each of the two valleys represents the opposite sublattice occupancy. Since the pseudo-Zeeman term is $\propto \bsigma\cdot\bB$, a lower energy state is achieved when the pseudo-spin is (anti)aligned with the pseudo-magnetic field in each valley. As the pseudo field changes sign in each valley, the pseudo-spins also align in opposite directions. This adjustment results into a charge redistribution (or charge imbalance) between the two sublattices and  manifests itself as a SSB in images of charge distributions.

In contrast, a real magnetic field enters the Dirac equation via a vector potential with the same sign in both valleys, causing pseudo-Zeeman terms that result in the alignment of the pseudo-spins in the same direction in each valley. As a consequence, in one valley one sublattice is more occupied but in the other valley it is the other one. Therefore, a real magnetic can not produce a true polarization of the pseudo-spin as the effect vanishes when the contributions from both valleys are added. SSB clearly demonstrates that a strain-induced pseudo-magnetic field is fundamentally different from a real magnetic field.

\begin{figure}[th]
	\centerline{\includegraphics[width=2.5in]{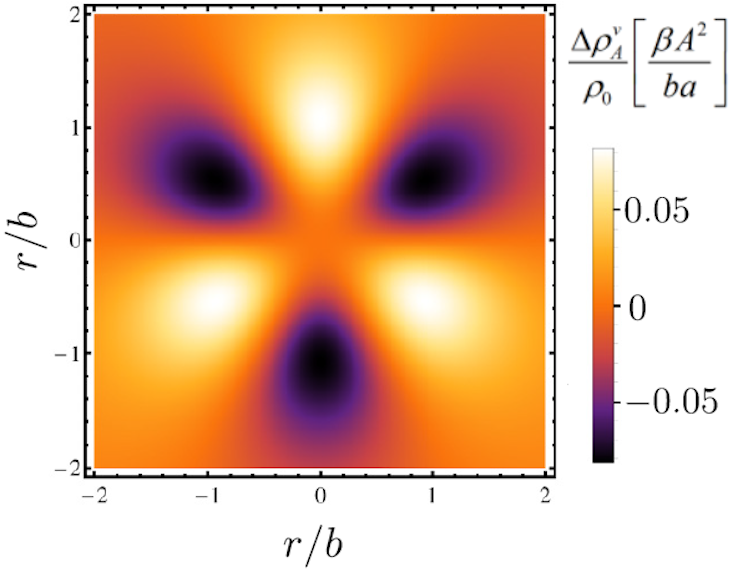}}
	\vspace*{8pt}
	\caption{Spatial profile of the change of LDOS for sublattice A in the presence of a Gaussian bump. Bright (dark) spots indicate an increase (decrease) of LDOS compared to pristine graphene. For sublattice B, the effect is exactly opposite. Figure adapted from Ref.~\citen{SSBMartin} with permission. \label{Fig:SSB_STM_Martin}}
\end{figure}

To visualize the SSB effect, one can analyze the LDOS in the deformed region. Using the Born approximation, analytic expressions for the change in the LDOS can be obtained when the strain intensity is small \cite{SSBMartin,MyPaperKondo,economou}. The change of LDOS exhibits opposite signs in the two sublattices, i.e. $\Delta\rho^v_A=-\Delta\rho^v_B$, where the subscripts label the sublattices and the superscript indicates that the effect is due to the pseudo-vector potential. Fig.~\ref{Fig:SSB_STM_Martin} presents typical results of $\Delta\rho^v_A$ normalized with respect to the pristine value $\rho_0$ as function of location in the deformed area in the presence of a Gaussian bump (Fig.~\ref{Fig:deformations}(a)). For a small bump that satisfies $kb\ll1$, where $k$ is the magnitude of the wave vector, the result is given by the following analytic expression\cite{SSBMartin}
\begin{equation}
	\frac{\Delta\rho^v_A}{\rho_0}=-\frac{\Delta\rho^v_B}{\rho_0}=-\frac{\beta h_0^2}{ab}\sin3\theta \,f\left(\frac{r}{b}\right),
\end{equation}
with $f(x)=\left[1-e^{-2x^2}\left(1+2x^2+2x^4\right)\right]/\left(4x^3\right)$.
One can clearly notice that $\Delta\rho^v_A$ exhibits a similar flower pattern as the pseudo-magnetic field (Fig.~\ref{Fig:B_and_Phi}(a)), which has three-fold symmetry with alternating positive and negative regions. $\Delta\rho^v_A$ and the pseudo-magnetic field share similar profiles is in general true for any deformation (see Ref.~\citen{ZhaiDissertation} for results corresponding a Gaussian fold), reflecting the fact that $\Delta\rho^v_A$ originates from the pseudo-Zeeman effect. 

\begin{figure}[th]
	\centerline{\includegraphics[width=3.5in]{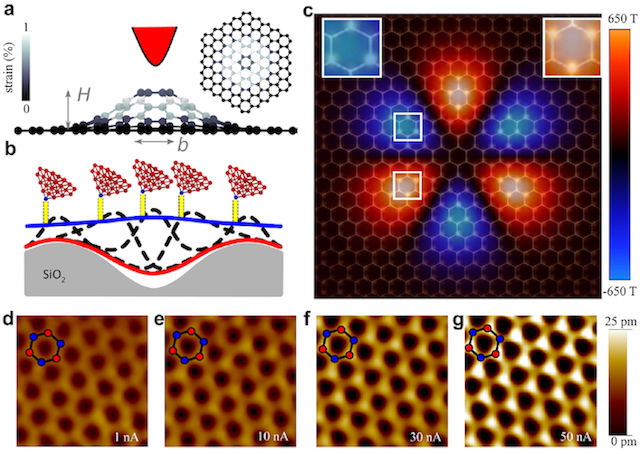}}
	\vspace*{8pt}
	\caption{(a, b) Schematics of STM tip lifted graphene bubbles. (c) Numerical results showing the SSB. (d--g) By increasing the tunneling current the bubble is gradually formed, and the transition of honeycomb to triangular lattice structure is observed. Figure reproduced from Ref.~\citen{MarkusNanoLett2017}  with permission. \label{Fig:SSB_STM_Alex}}
\end{figure}

Due to the opposite changes of LDOS in the two sublattices, images of atoms in different sublattices will exhibit a contrast in local probe measurements, rendering a triangular lattice structure instead of the standard honeycomb. This SSB was already observed several years ago on graphene with wrinkles, and STM tip lifted graphene bubbles but attributed to different origins~\cite{SSBNanoLett2009,MarkusNanoLett2010}. Very recently, systematic quantitative evaluations of the SSB effect were performed using STM on graphene bubbles that confirmed predictions based on a pseudo-spin polarization due to a strain-induced pseudo-magnetic field (Fig.~\ref{Fig:SSB_STM_Alex})~\cite{MarkusNanoLett2017}.

The charge redistribution, strongly dependent on the profile of the underlying pseudo-magnetic field, might be utilized to determine the strain field of different deformations. Furthermore, it might also facilitate the observation of phenomena closely related to local variations of charge densities~\cite{Hewson1997}, as is the case of the Kondo effect. The observation of Kondo effect in graphene has been challenging and existing results remain controversial for various reasons. Deformed graphene structures exhibiting SSB can function as simple platforms for observing a sublattice resolved Kondo effect pattern that is unique to the material. We refer the reader to Ref.~\citen{MyPaperKondo} for results of a systematic study of $\Delta\rho^v_{A,B}$ as function of energy and the accompanying Kondo temperature patterns.

\subsection{Confinement}\label{Sect:Confinement}
The above discussions indicate that strain induced pseudo-magnetic field can be utilized to manipulate the pseudo-spin degree of freedom and achieve charge accumulation on the sublattice level. Such behavior involves charge transfer from one sublattice to the other within the deformed area, while the effect is negligible outside it. These observations raise the question of whether deformations can be utilized to effectively confine particles without the use of external voltages, a mission proven to be quite challenging due to the gapless dispersion and the Klein tunneling effect\cite{KleinTunneling,KleinTunnelingExpt}. In fact, previous theoretical studies have seen signatures of confinement in graphene bumps, in the form of either circulating currents inside the deformed region, or dips in the conductance plateaus.\cite{Confinement_Nanoscale,RamonBump2014} In this section, we discuss the confinement effect using the fold geometry that can be represented by an effective one-dimensional model and thus allows analytical treatment in some limiting parameter regimes.

\begin{figure}[ht]
	\centerline{\includegraphics[width=5in]{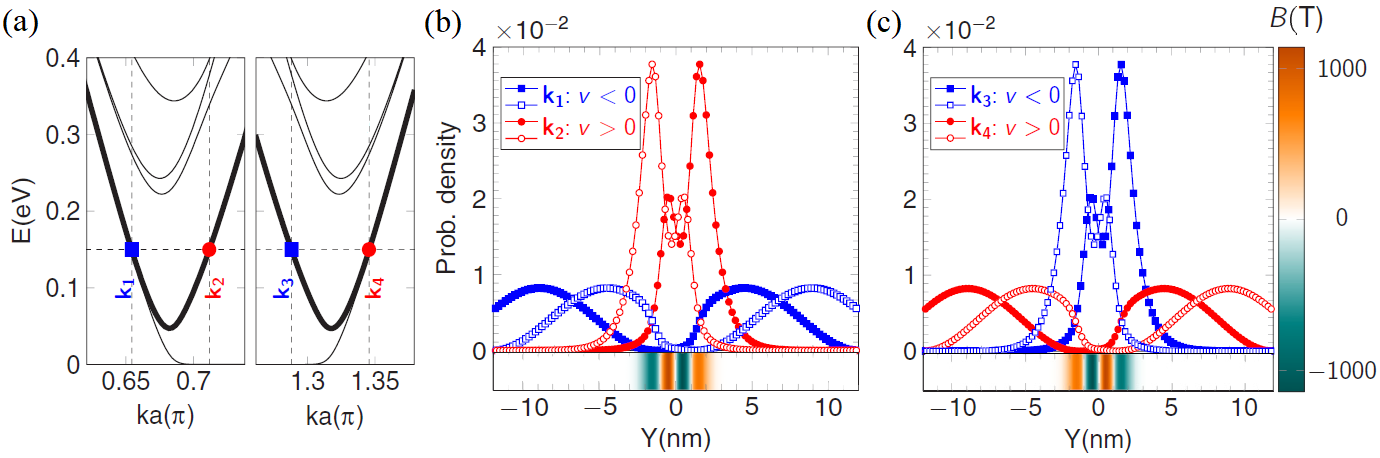}}
	\vspace*{8pt}
	\caption{(a) Low energy band structure for a graphene Gaussian fold: left $K$ and right $K'$, respectively. (b-c) Probability densities for states at energy $E = 0.15$ eV as indicated by the symbols in panel (a). Blue (red) curves correspond to states $k_1$ and $k_3$ ($k_2$ and $k_4$) with negative (positive) group velocity. Filled (empty) symbols indicate sublattice A (B). Color scale indicates magnitude of pseudo-magnetic field. Figure adapted from Ref.~\citen{RamonFold} with permission. \label{Fig:Ramon_fold}}
\end{figure}

Fig.~\ref{Fig:Ramon_fold}(a) shows results for the low energy dispersion around the Dirac points, for a fold extended along the zigzag crystalline orientation, as reported in Ref.~\citen{RamonFold}. Panels (b) and (c) show the probability distribution for selected states at energy $E=0.15$ eV, indicated by dot and square symbols in panel (a). One can clearly see that in general the probability distributions on the two sublattices are different, consistent with the SSB effect. 
Because of the zigzag nanoribbon geometry, the probability distribution does not completely vanish near the edges due to the contribution of the edge states. However, one can still notice that the majority of the wave functions lie in the fold region, especially for states labeled by $k_2$ and $k_3$. These results clearly imply that deformations do confine particles in graphene. 

To describe the origin of this confinement, we turn back to Eq.~(\ref{Eq:foldfields}) where the pseudo-vector potential of a Gaussian fold exhibits a double barrier structure due to the function $g(y/b)$. As a simplification, the smooth potential can be approximated by a double square barrier geometry (Fig.~\ref{Fig:Double_barrier}(b)). Interestingly, the basic ingredients that give rise to confined states are already captured with a single square barrier model: $\bA=(A_x=\text{const.},0)$ in region $y\in[0,b]$, analyzed below.

The main feature of confined states is that their wave functions are localized in the region where the potential resides. Considering a scattering approach, a free particle state with wave vector $\bk$ and energy $E = \hbar v_F k$ ($k = |\bk|$), has the momentum components, $k_x$ and $k_y$, constrained by $k_y=\sqrt{k^2-k_x^2}$. Notice that   $k_y$ can take imaginary values for $|k_x|> k$. Under such a circumstance, the wave functions decay exponentially outside the barrier. To see this explicitly, one can define $\kappa=\sqrt{k_x^2-k^2}$ for $|k_x|> k$, then $k_y=i\kappa$, and the wave function on either side of the barrier has the form
\begin{equation}
\begin{aligned}
\Psi_{L}&=
C_1
\begin{pmatrix}
1\\\frac{k_x+\kappa}{k}
\end{pmatrix}
e^{ik_xx}e^{\kappa y}\\
\Psi_{R}&=
C_2
\begin{pmatrix}
1\\\frac{k_x-\kappa}{k}
\end{pmatrix}
e^{ik_xx}e^{-\kappa y}
\end{aligned}.\label{Eq:BoundStatePsiOutsideBarrier}
\end{equation}
where $C_1, C_2$ are appropriate normalization constants. These wave functions vanish at $|y|\rightarrow\infty$, and correspond to confined states.
The wave function inside the potential barrier region has the form
\begin{equation}
\Psi=C_1
\begin{pmatrix}
1\\\frac{q_x+iq_y}{k}
\end{pmatrix}
e^{iq_yy}e^{ik_xx}
+
C_2
\begin{pmatrix}
1\\\frac{q_x-iq_y}{k}
\end{pmatrix}
e^{-iq_yy}e^{ik_xx},\label{Eq:Psi_extended}
\end{equation}
where $(q_x=k_x\pm eA_x/\hbar, q_y)$ are the components of the wave vector in the barrier and $\pm$ corresponds to each valley. Notice that in the case of a double barrier potential, similar wave functions also exist, and a linear combination of solutions in the form of Eq.~(\ref{Eq:BoundStatePsiOutsideBarrier}) produces the solution in the well between the barriers.

The two exponential functions are finite inside the region, and decay exponentially when moving away from the potential. These states are thus bounded in the $\hat{y}$ direction (across the potential), while they can freely propagate along the $\hat{x}$ direction, behaving like wave-guided modes. These results have an immediate interpretation in terms of new `edge states' that are created by the fold.  

\begin{figure}[ht]
	\centerline{\includegraphics[width=4in]{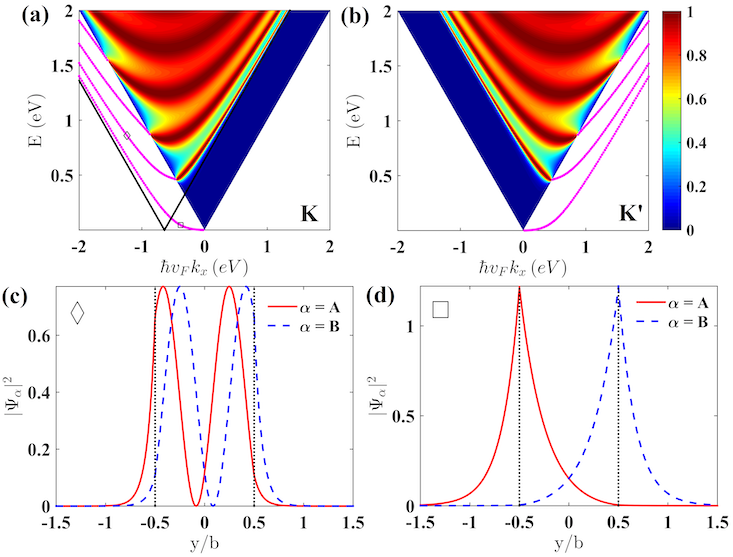}}
	\vspace*{8pt}
	\caption{(a, b) Confined state dispersion (magenta curves) as function of energy $E$ and $\hbar v_F k_x$ for valley $K$ and $K'$ in the presence of a square pseudo-vector potential barrier. (c, d) Probability density and of two typical confined states for valley $K$ as indicated by the diamond and square symbols in (a). The dotted vertical lines indicate the edges of the potential barrier. Discontinuities in the slopes of the curves at the boundaries are artificially caused by the sharp edges. Height of the barrier: $|v_FA_x|=0.64$ eV, width of the barrier: $b=5$ nm. Figure reproduced from Ref.~\citen{ZhaiDissertation}. \label{Fig:Single_A_barrier_transmission_and_bound_states}}
\end{figure}

By matching the wave functions in different regions, one can obtain the confined state dispersion. The magenta curves in Fig.~\ref{Fig:Single_A_barrier_transmission_and_bound_states}(a, b) represent the confined state dispersion for a square pseudo-vector potential barrier. The filled triangles with color map ranging from blue to red represent the Dirac cones (corresponding to $|k_x|\le k$). These states have wave functions extended throughout the space and electrons associated with these states have a finite transmission probability (color coding). We focus on such behavior in Sect.~\ref{Sect:Valley_filter_fold}. The $K$ and $K'$ valleys are related by time-reversal symmetry, as shown by the mirror symmetric structures in  Fig.~\ref{Fig:Single_A_barrier_transmission_and_bound_states}(a, b). Due to the fact that $|k_x|> k$ for confined states, discrete bands only exist outside the Dirac cone. The black lines, corresponding to the shifted Dirac cones defined by $q_y=0$ inside the potential barrier, divide the confined states into two groups depending on whether $q_y$ is real or imaginary.\cite{Confinement_A_barrier_PRL2009,Confinement_A_barrier_RMP2010} In the region closer to the Dirac point (region I), $q_y$ is imaginary, and the corresponding wave function has exponential dependence that signals localization near the two edges. The corresponding energy dispersion can be considered as a precursor of Landau levels.\cite{Marc2018} In the region with higher energies (region II), $q_y$ is real, and the wave function inside the barrier has a sine/cosine oscillatory profile.

It is also interesting to look at the scaling of the energy dispersion, which might be of interest in e.g. STM measurements. At $k_x\approx\pm eA_x/\hbar$, i.e. near the center of the shifted Dirac cones, the energy can be approximated as $E_n\approx (n+1/2)\hbar v_F/b$.\cite{Confinement_A_barrier_PRL2009} This implies that the confined state bands are roughly equally spaced and inversely proportional to the width of the barrier. A recent STM experiment on a graphene wrinkle has revealed dI/dV peaks with equal spacing, which decreases with an increasing wrinkle width, consistent with the confinement effect.\cite{FoldNatCommun2015} In contrast, Landau levels produced by a strong uniform (pseudo)magnetic field, scale as $E_n\propto\sqrt{n}$.\cite{GoerbigRMP,Crommie} Such scaling is obviously different from the equal spacing produced by confinement. The competition between these two scaling regimes can be cast in terms of the ratio $b/l_B$, that highlights the two different confinement mechanisms.
A limited number of pseudo-Landau levels may develop near the Dirac point (e.g. region I in Fig.~\ref{Fig:Single_A_barrier_transmission_and_bound_states}(a)) for $b \gg l_B$.\cite{LandauLevelScalingPRB2008,LandauLevelScalingPRB2013} In such a situation, the extension of the region $b$ with a pseudo-magnetic field, is larger than the magnetic length $l_B=\sqrt{\hbar/(eB)}$ associated with the strength of the pseudo-magnetic field $B$. The number of Landau levels that can be observed in STM experiments can be roughly estimated by this ratio. Usually experiments involve complicated magnetic field profiles, so the measured scaling of energy levels might be a mixture of these two scalings (strain induced scalar potential can also affect Landau levels, see e.g. Ref.~\citen{LandauLevelScalingPRB2017}), thus special care should be given to explain its origin.

In order to visualize the confinement effect, one can plot the probability distribution of these confined states. Fig.~\ref{Fig:Single_A_barrier_transmission_and_bound_states}(c, d) show representative probability densities in two different regions indicated by the diamond and square symbols in panel (a). Here the discontinuity in the slope is caused by the sharp edges of the potential barrier (Note that  probability densities do not necessarily have smooth derivatives for Dirac fermions where only continuity of wave function is required.~\cite{RainisFold}.). The oscillatory behavior inside the barrier (with boundaries indicated by the vertical dotted lines) for the state labeled by the diamond symbol corresponds to a real $q_y$, while the localization near the potential edges for the state labeled by the square symbol, corresponds to an imaginary value for $q_y$.

\begin{figure}[ht]
	\centerline{\includegraphics[width=5in]{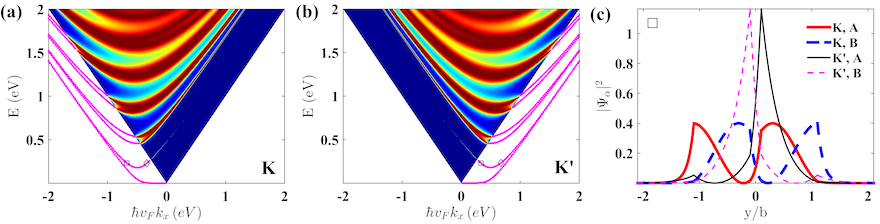}}
	\vspace*{8pt}
	\caption{Confined state dispersion for valley $K$ (a) and $K'$ (b) in the presence of double pseudo-vector potential barriers. The probability density of states from the two valleys with similar energy and group velocity as indicated by the squares symbols are shown in (c). The thick (thin) curves are for the $K$ ($K'$) valley. States labeled by diamonds in (a) are related to those labeled by squares via time-reversal symmetry. The probability density distributions have the same profile as those in (c) but with the exchange of valley index. Height of the barrier: $|v_FA_x|=0.64$ eV, width of the barriers: $b=5$ nm, separation of the barriers: $0.2b=1$ nm. Figure reproduced from Ref.~\citen{ZhaiDissertation}. \label{Fig:Double_A_barrier_transmission_and_bound_states}}
\end{figure} 

In order to compare with Fig.~\ref{Fig:Ramon_fold} directly, we present results for the double square barrier structure in Fig.~\ref{Fig:Double_A_barrier_transmission_and_bound_states} that shows confined state dispersions (magenta curves) in the presence of a double pseudo-vector potential barrier separately for each valley. Due to the presence of the second barrier, the confined state bands are broken into pairs at the boundaries of the Dirac cones and merge again for large $|k_x|$ and high $E$. One may notice that the low energy dispersion is very similar to that in Fig.~\ref{Fig:Ramon_fold}(a) obtained from tight-binding calculations.
Similar to Fig.~\ref{Fig:Ramon_fold}, we select states from the two valleys with similar energy and band group velocity as indicated by the square symbols on the bands (Fig.~\ref{Fig:Double_A_barrier_transmission_and_bound_states}(a, b)). A plot of the corresponding probability density distribution is shown in panel (c), with thick (thin) curves for valley $K$ ($K'$) and solid (dashed) curves labeling the $A$ ($B$) sublattice contribution. These results are consistent with those in Fig.~\ref{Fig:Ramon_fold}(b, c), i.e. the spatial occupations of the two states are distinct, with the central (side) part of the double barrier structure dominated by the $K'$ ($K$) valley. For the states marked by diamonds symbols in panels (a, b), the profile of the probability density distribution remains the same after the exchange of valleys.

\begin{figure}[ht]
	\centerline{\includegraphics[width=3.5in]{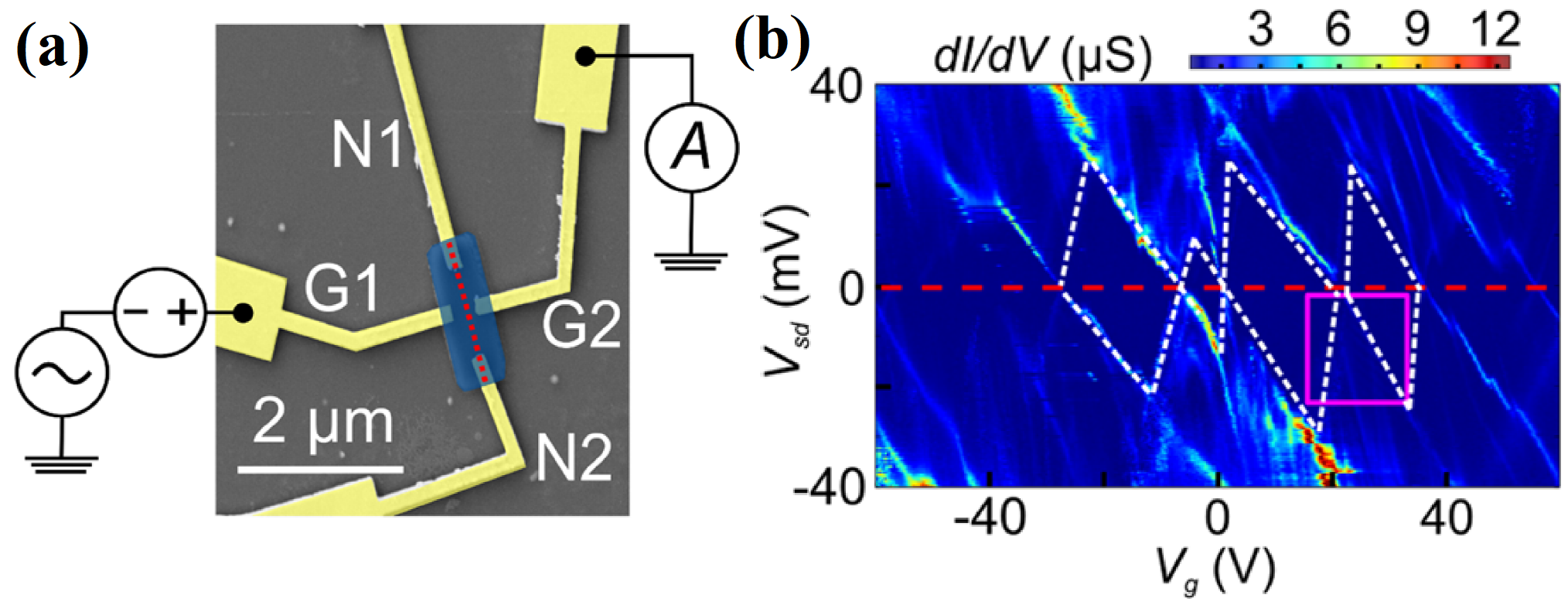}}
	\vspace*{8pt}
	\caption{(a) Colored image of the transport device composed of graphene (blue), four contacts (yellow), and a fold (red dashed line). (b) Representative conductance measurement across the fold, where Coulomb blockade diamonds can be identified. Figure adapted from Ref.~\citen{Marc2018} with permission. \label{Fig:Marc_expt}}
\end{figure}

The confinement effect discussed above is in good agreement with recent experiments showing Coulomb blockade diamonds (Fig.~\ref{Fig:Marc_expt}(b))-- signatures of single electron charging and confinement-- in transport conductance measurements when low energy currents are injected across graphene folds (Fig.~\ref{Fig:Marc_expt}(a))~\cite{Marc2018}. Coulomb blockade physics is consistent with theoretical predictions that extended states are removed from the low energy spectrum (see Sect.~\ref{Sect:Valley_filter_fold} in the following), and confined states allow transport via electron tunneling. 

\begin{figure}[ht]
	\centerline{\includegraphics[width=3in]{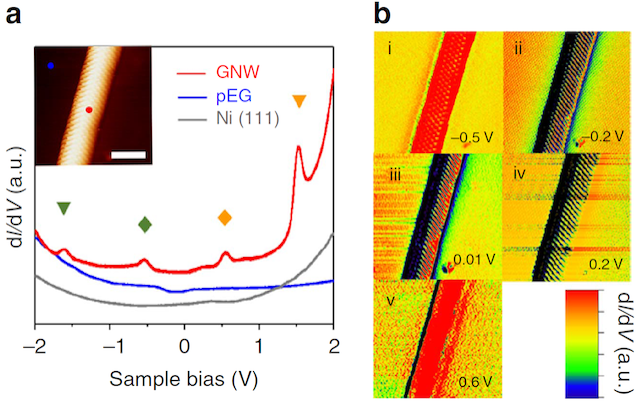}}
	\vspace*{8pt}
	\caption{(a) dI/dV spectra on (grey line) Ni(111), (blue line) flat graphene, and (red line) graphene nanowrinkle as marked in inset STM image. Scale bar, 2 nm. (b) dI/dV mapping images of graphene nanowrinkle obtained at various bias voltages. Figure reproduced from Ref.~\citen{FoldNatCommun2015} with permission. \label{Fig:Confinement_natcommun}}
\end{figure}

Another experimental example of confinement in deformed graphene involves graphene nanowrinkles formed by depositing graphene on Ni(111) (Fig.~\ref{Fig:Confinement_natcommun}).\cite{FoldNatCommun2015} Equally spaced dI/dV peaks were observed (Fig.~\ref{Fig:Confinement_natcommun}(a)), whose separation was found to be inversely proportional to the width of the wrinkles. The dI/dV mapping shown in Fig.~\ref{Fig:Confinement_natcommun}(b) reveal a confinement effect where the weights of the states are localized in the wrinkle.

\subsection{Particle-hole asymmetry due to scalar potential}
\begin{figure}[ht]
	\centerline{\includegraphics[width=2in]{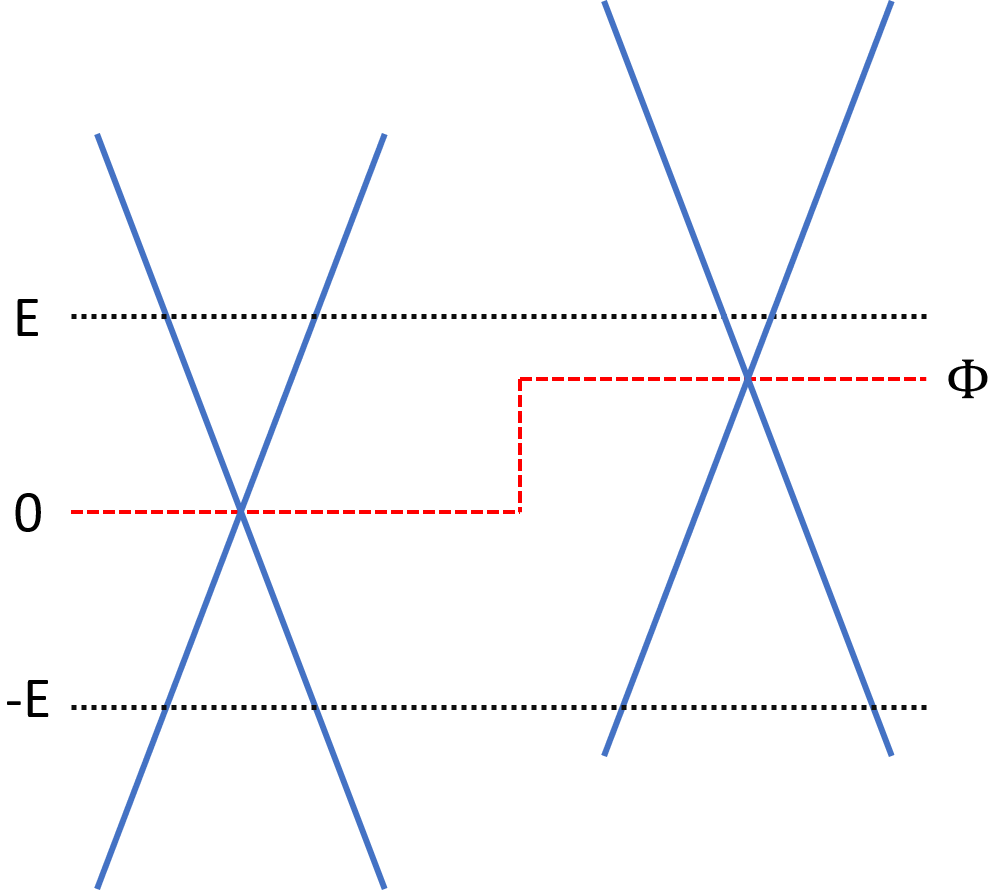}}
	\vspace*{8pt}
	\caption{Schematics of the effect of the scalar potential $\Phi$ (red dashed lines), which shifts the unperturbed Dirac cone (left) by an amount $\Phi$ in energy (right). Figure reproduced from Ref.~\citen{ZhaiDissertation}. \label{Fig:schematics_particle_hole_asymmetry}}
	
\end{figure}

Now let's focus on the effect of the scalar potential due to strain. The Schr\"odinger equation reads $(v_F\bsigma\cdot\textbf{p}+\Phi)\Psi=E\Psi$, or equivalently, $v_F\bsigma\cdot\textbf{p}\Psi=(E-\Phi)\Psi$. Therefore, the effect of the scalar potential is to shift the energy, i.e. $E\rightarrow E-\Phi$, as shown schematically in Fig.~\ref{Fig:schematics_particle_hole_asymmetry}. Without the scalar potential, the number of states at energy $\pm E$ are the same (equal widths of black dotted line cuts at $\pm E$ on the cone), i.e. the system exhibits particle-hole symmetry. In the presence of the scalar potential $\Phi$, due to the fact that the Dirac cone is shifted in energy, the number of states are different at energies $\pm E$, and the particle-hole symmetry is broken. 

Strictly speaking, this simple picture is only valid for a constant scalar potential spanning the whole sample area, which does not mix states with different momenta. For a scalar potential that is localized in a finite region with typical width $b$ (e.g. Fig.~\ref{Fig:B_and_Phi}), the change in the LDOS and the particle-hole symmetry breaking is more complicated. However, in the case of high energies ($E\gg E_b=\hbar v_F/b$), the wavelengths of the particles are much smaller than $b$. Thus in the high energy limit the potential appears to be constant, and the above simple picture remains valid.

\begin{figure}[ht]
	\centerline{\includegraphics[width=4in]{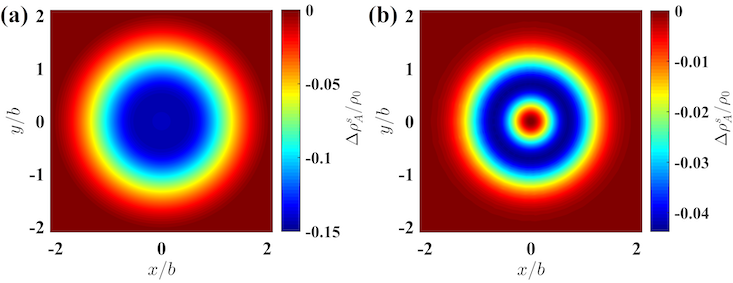}}
	\vspace*{8pt}
	\caption{$\Delta\rho^{s}_{K,A}$ as function of position at  $E=0.1$ eV (a) and $E=0.4$ eV (b) in the case of a Gaussian bump. Note that $\Delta\rho^{s}_{K,A}$ is normalized with respect to the LDOS for pristine graphene $\rho_0$. Figure reproduced from Ref.~\citen{ZhaiDissertation}. \label{Fig:scalar_Delta_LDOS_vs_position_bubble}}
\end{figure}

Although the scalar potential, which does not couple to the pseudo-spin, respects the sublattice symmetry, it affects charge distribution as well. Fig.~\ref{Fig:scalar_Delta_LDOS_vs_position_bubble} presents two typical results of the change of LDOS $\Delta\rho^s_A$ as function of position due to the scalar potential (indicated by the superscript) for $A$ sublattice (indicated by the subscript) in the presence of a Gaussian bump (results corresponding to the case of a fold can be found in Ref.~\citen{ZhaiDissertation}). The $B$ sublattice shares identical results. Due to the angular isotropy of the scalar potential, $\Delta\rho^{s}_A$ exhibits circular symmetry, however, it varies in the radial direction, with the largest changes in the area of the bump. Especially, at large energies $E\gg E_b$, $\Delta\rho_A^{s}$ shares the same structure as the scalar potential (Fig.~\ref{Fig:scalar_Delta_LDOS_vs_position_bubble}(b)). To be more specific, $\Delta\rho_A^{s}(\br,E\gg E_b)\rightarrow -\Phi(\br)/\left(2\pi\hbar^2v_F^2\right)$, consistent with the claim that in the high energy limit the effect of the scalar potential is equivalent to a shift in the energy $E\rightarrow E-\Phi$.

Fig.~\ref{Fig:scalar_Delta_LDOS_vs_position_bubble} clearly shows that the scalar potential reduces the LDOS in the deformed area for negatively doped samples; while the LDOS will be enhanced for positively doped samples, i.e. it can also contribute to charge confinement (Note that the conclusion is obtained by assuming $g_s=3$ eV.). For more results regarding $\Delta\rho^{s}_A$ as function of energy for the case of both Gaussian bump and fold geometries, we refer the reader to Ref.~\citen{MyPaperKondo}.

\section{Valley filtering}\label{Sect:ValleyFilter}
In transport experiments, the incident current is composed of equal amount of electrons from both valleys. Because strain induces a pseudo-magnetic field that exhibits opposite signs in the two valleys, it is expected that valley polarized currents might be obtained by passing current through a deformation as electrons from different valleys undergo distinct trajectories. Such valley polarized current carrying one specific valley index could be used for valleytronic device applications\cite{ValleytronicsReview}.

In this section we analyze the transport properties of graphene in the presence of deformations -as shown in Fig.~\ref{Fig:deformations}- in terms of the resulting valley polarization of currents incident in the deformed region.\cite{Settnes,PeetersBump,MyPaperValleyFilter,RamonBump2018,SzpakBump2018,EnriqueBubble2017} The purpose is to characterize the effectiveness of these structures as valley filters in terms of parameters with experimental relevance in available setups. Various other graphene valley filter proposals have also been suggested\cite{ValleyNatPhys2007,ValleyPRL2007,TrigonalWarpingPRL2008,TrigonalWarpingJPhysCondensMatter,TrigonalWarpingLight,LightBilayer,ValleyStrainPRL,ValleyStrainMagnetAPL,ValleyStrainMagnetPRB,ValleyStrainACvoltagePRL,TransmissionOutofPlaneDeformationPRL2008,ValleyMassBarrierPRB,ValleyLineDefectPRL2011,ValleyLineDefectPRB2014,ValleytronicsNatPhysBilayerGraphene,ValleytronicsNatPhysBilayerGraphene2,PeetersValleyFilterPRB2016,RamonFold,ValleyFilterCorbino}, one can refer to Ref.~\citen{MyPaperValleyFilter} for a brief overview.

\subsection{Bubble geometry}

\begin{figure}[ht]
	\centerline{\includegraphics[width=5in]{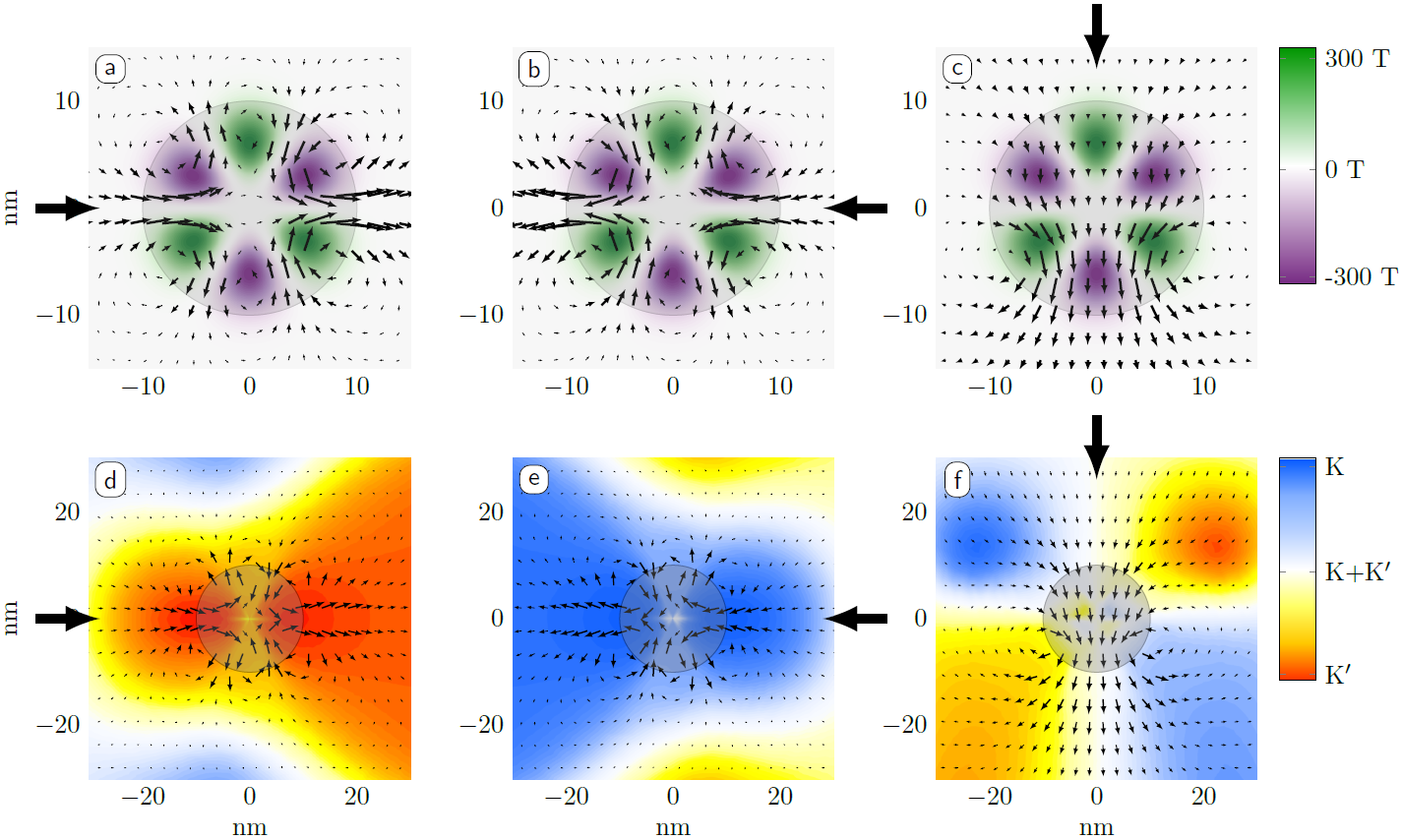}}
	\vspace*{8pt}
	\caption{Valley polarization with different incident directions for the Gaussian bump. (a-c) An incident wave with $E = 0.01t$ from the left, right and top as indicated by the large black arrows. The local currents are indicated by the arrow map, and the colormap illustrates the pseudo-magnetic field distribution. (d-f) Real space map of the relative occupation of $K$ and $K'$ corresponding to the different incident directions. Figure reproduced from Ref.~\citen{Settnes} with permission. \label{Fig:Settnes_valley_filter}}
\end{figure}

From the fact that electrons from the two valleys experience opposite magnetic fields, Settnes et al showed, with tight-binding calculations, that graphene nanobubbles can be utilized to achieve strong valley polarization (Fig.~\ref{Fig:Settnes_valley_filter}).\cite{Settnes} The effect of strain was incorporated via changes in the hopping energy through carbon-carbon distance modifications (Eq.~(\ref{Eq:t_modified})). Similar calculations were performed later by adding contacts into the system.\cite{SzpakBump2018,PeetersBump} The main findings are: (i) The current near the bubble can be enhanced and exhibits valley polarization that depends on angular direction (Fig.~\ref{Fig:Settnes_valley_filter}(a, d)); (ii) The valley polarization originates from the effect of the underlying pseudo-magnetic field and strongly depends on the incident direction (Fig.~\ref{Fig:Settnes_valley_filter}(b, e, c, f)). Especially, it can switch sign if the profile of the pseudo-magnetic field changes sign with respect to the incident direction (Fig.~\ref{Fig:Settnes_valley_filter}(d) vs (e)). 

\subsubsection{Scattering model of valley polarization}
All these findings can be understood analytically via the scattering theory in the continuum formalism. To compare with tight-binding results, we focus on the effect of the pseudo-magnetic field first. A recent study\cite{MyPaperValleyFilter} has shown that electrons coming from the two valleys exhibit different scattering cross sections in the second order Born approximation: 
\begin{equation}
\sigma^{(2),\tau}=\frac{k}{2\pi\hbar^2v_F^2}\left|\tau V_{\bk,\bk'}+(VGV)_{\bk,\bk'}\right|^2\approx\sigma^{(1)}+\tau\frac{1}{2}\Delta,\label{eq:sigmasforKK'}
\end{equation}
where $\tau=\pm$ labels the two valleys, the numeric subscripts label the order of the approximation, $V$ is the potential due to the pseudo-magnetic field in $K$ valley, $G$ is the retarded Green's function of pristine graphene, the subscripts $\bk'$ and $\bk$ indicate incident and outgoing wave vectors, and 
\begin{equation}
\begin{aligned}
\Delta=\sigma^{(2),+}-\sigma^{(2),-}=\frac{2k}{\pi\hbar^2v_F^2}\text{Re}\left[V_{\bk,\bk'}\cdot (VGV)_{\bk,\bk'}\right]
\end{aligned}
\end{equation}
is the difference between the differential cross sections of the two valleys. Second order terms are necessary to observe valley polarization, i.e. different scattering cross sections in the two valleys, giving a scaling for $\Delta$ with the strain intensity of $\eta^6$. 

\begin{figure}[ht]
	\centerline{\includegraphics[width=3in]{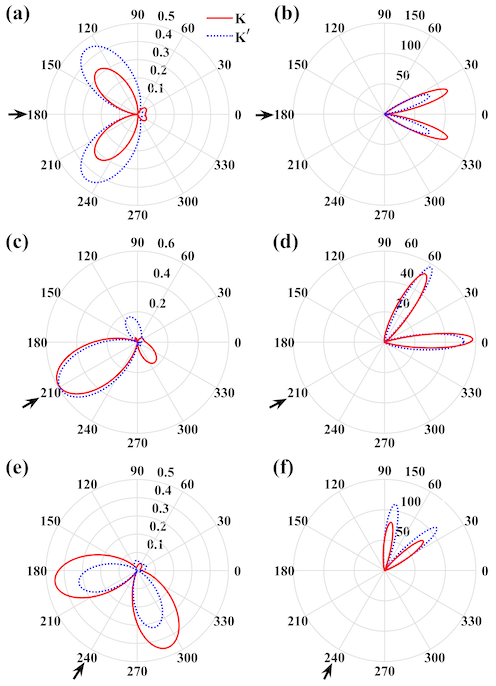}}
	\vspace*{8pt}
	\caption{Polar plots of the differential cross section (in units of \AA) for $K$ (red solid) and $K'$ (blue dotted) valleys with different incident angles indicated by the arrows and total cross section as function of incident angles. (a-b) $\theta'=0^\circ$, (c-d) $\theta'=30^\circ$, (e-f) $\theta'=60^\circ$. Left column corresponds to energy $E=20$ meV, while right column corresponds to energy $E=300$ meV. Other parameters: $b=15$ nm, $\eta=0.1$. Figure adapted from Ref.~\citen{MyPaperValleyFilter} with permission.\label{cross_section_vector_only}}
\end{figure}

Fig.~\ref{cross_section_vector_only} presents results of the scattering cross section in the low and high energy regimes in the left and right columns respectively, for an incident electron from three representative directions. The figures show that the degree of valley polarization is highly dependent on the incident direction, with maximum polarization occurring for the zigzag crystalline direction (even multiples of $30^\circ$) in both regimes. However, for particles incoming along the armchair directions (odd multiples of $30^\circ$), the valley polarization is the weakest. Notice that the polarization is reversed as the incident angle changes by $60^\circ$, e.g. the first and third rows of Fig.~\ref{cross_section_vector_only}, consistent with the changes in the underlying pseudo-magnetic field pattern as the rotation is carried out. This suggests that the magnitude of the polarization can be switched by properly controlling the incident direction.
All these observations and the symmetric or anti-symmetric profiles depending on the incident direction are consistent with tight-binding results in Fig.~\ref{Fig:Settnes_valley_filter}. The profiles of the differential cross sections can be intuitively understood by considering the classical motion of electrons in a magnetic field.\cite{MyPaperValleyFilter} For instance, electrons with low (high) energy are more likely to be reflected (transmitted), thus backward (forward) scattering is observed. The symmetric or anti-symmetric profiles are also consistent with the underly pseudo-magnetic field distribution.
Overall, the filtering is more effective at high energies, giving bigger differential cross sections. Notice however that currents for both valleys coexist in the same spatial region, making the scattered currents only partially polarized.

\begin{figure}[ht]
	\centerline{\includegraphics[width=3in]{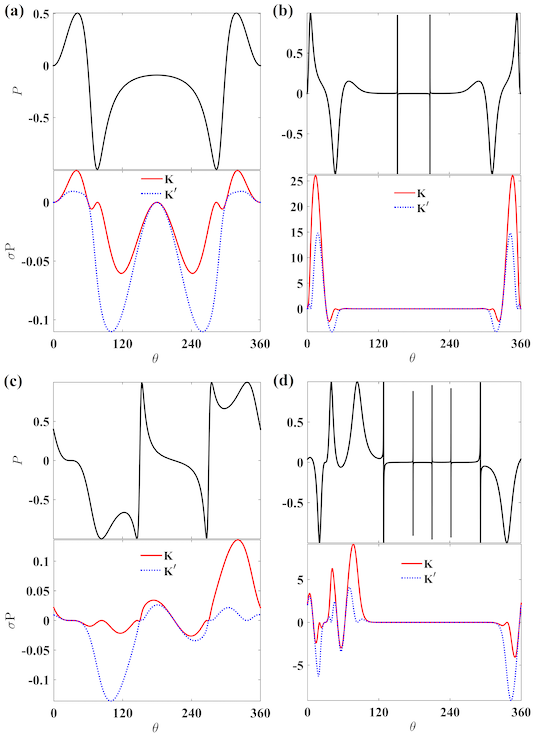}}
	\vspace*{8pt}
	\caption{Polarization $P$ (black), $\sigma P$ for $K$ (red solid) and $K'$ (blue dotted) valley corresponding to panels (a-d) in Fig.~\ref{cross_section_vector_only}. Figure reproduced from Ref.~\citen{MyPaperValleyFilter} with permission. \label{P_and_SigmaP}}
\end{figure}

In order to quantify the degree of valley polarization, one needs the angle dependent polarization coefficient $P$ defined as:
\begin{equation}
P=\frac{\sigma^{(2),+}-\sigma^{(2),-}}{\sigma^{(2),+}+\sigma^{(2),-}}\approx\frac{\Delta}{2\sigma^{(1)}}.
\end{equation}
This expression indicates that increased polarization can be achieved by larger strain intensities as $P \propto \eta^4$.
A large product $\sigma P$ is also preferable as both a measurable signal (large $\sigma$) and large polarization are desirable for a good valley filter.\cite{SzpakBump2018}

Fig.~\ref{P_and_SigmaP} shows the results corresponding to panels (a-d) of Fig.~\ref{cross_section_vector_only}. In the low energy regime (left column, black curves) large polarization values can be obtained in wide angular regions. However, the corresponding differential cross section has small magnitude, indication of weak scattering. Therefore, the product $\sigma P$, used to estimate the amount of detected current, is small (red solid and blue dotted curves). In the high energy regime (right column), the differential cross section reaches substantial values, however, the scattering events are confined to very narrow angular regions around the incident directions. Consequently, both $P$ and $\sigma P$ show narrow peaks and dips in the high energy regime. The narrow angular distribution and the close proximity to the incident direction impose serious difficulties for the detection of the scattered current: a detector (or contacts) with high angular resolution is required, and the incident current, usually fully unpolarized, is likely to overwhelm the weakly polarized currents.\cite{PeetersBump} 

\begin{figure}[ht]
	\centerline{\includegraphics[width=3in]{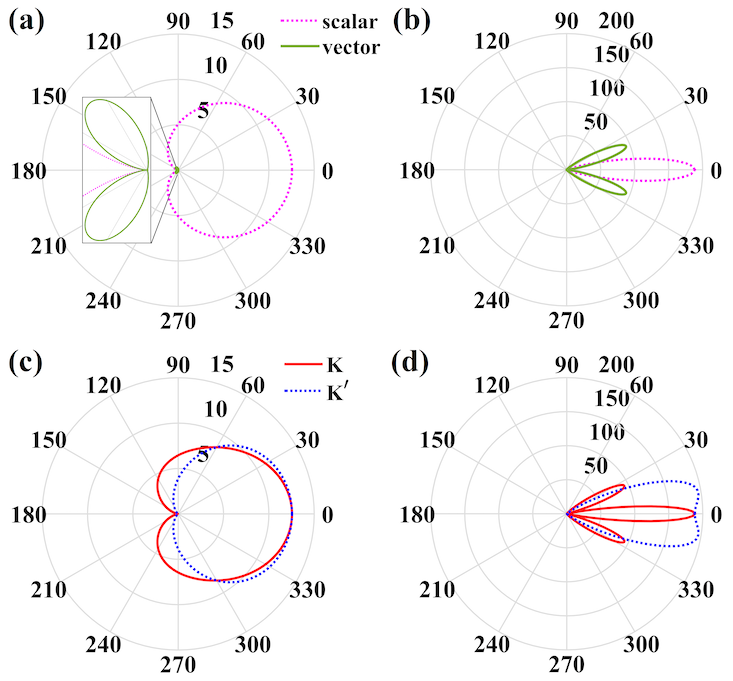}}
	\vspace*{8pt}
	\caption{Polar plots of the differential cross section (in units of \AA) for contributions of scalar (pink dotted) and pseudo-vector (green solid) potentials for valley $K$ and incident angle  $\theta'=0^\circ$. Panel (a) corresponds to energy $E=20$ meV. Inset shows a zoom-in to visualize the magnitude of the pseudo-vector potential contribution. Panel (b) corresponds to energy $E=300$ meV. Panels (c) and (d): differential cross section per valley at low and high energy respectively due to both potentials. Figure adapted from Ref.~\citen{MyPaperValleyFilter} with permission. \label{cross_section_with_scalar}}
\end{figure}

Effect of the scalar potential $\Phi$ on electron transport has been discarded up to this point. In tight-binding calculations, one need to consider next nearest-neighbor contributions in order to capture this effect,\cite{PeetersScalarPotential} thus it is rarely discussed in previous studies.\cite{Settnes,PeetersBump,SzpakBump2018} In the continuum formalism it is achieved by straightforwardly adding the scalar potential in the cross section formula.\cite{RamonBump2018,MyPaperValleyFilter} By itself, the scalar potential cannot give rise to valley filtered currents due to its valley-independent nature. However, when combined with the pseudo-vector potential $\tau V$, it renders a total scattering potential equals to $\Phi+\tau V$, which is different in the two valleys due to the sign change of $V$. Because the differential cross section is given by the form factor squared, it is clear from this expression that the two valleys will display different behavior even at the level of the first order Born approximation. Such valley polarization due to the addition of the scalar potential within first order Born approximation has also been addressed in a recent study.\cite{RamonBump2018}

Results of similar scattering calculations at first order, including the scalar potential are shown in Fig.~\ref{cross_section_with_scalar}. Panel (a), in the low energy regime, shows a much larger differential scattering cross section due to the scalar field than from the pseudo-vector potential, while comparable values are obtained for the high energy regime as shown in panel (b). Panels (c) and (d) present results from the total scattering potential, i.e., $\Phi + \tau V$ for the two valleys at low and high energy regimes respectively. In both cases, the structure still exhibits the valley filtering capability. However, the details of the valley polarization effect are dramatically different from those in the absence of the scalar potential (Fig.~\ref{cross_section_vector_only}).

\subsubsection{Drawbacks of graphene bumps for valley polarization}
We have seen from the scattering results that the scattered currents are usually composed of electrons from both valleys (i.e. partially polarized) and exhibit narrow angular distribution (Figs.~\ref{cross_section_vector_only} and \ref{P_and_SigmaP}). These features certainly undermine the utility of graphene bumps
as valley filters. In this section we discuss the drawbacks of graphene bumps. 

\begin{figure}[ht]
	\centerline{\includegraphics[width=5in]{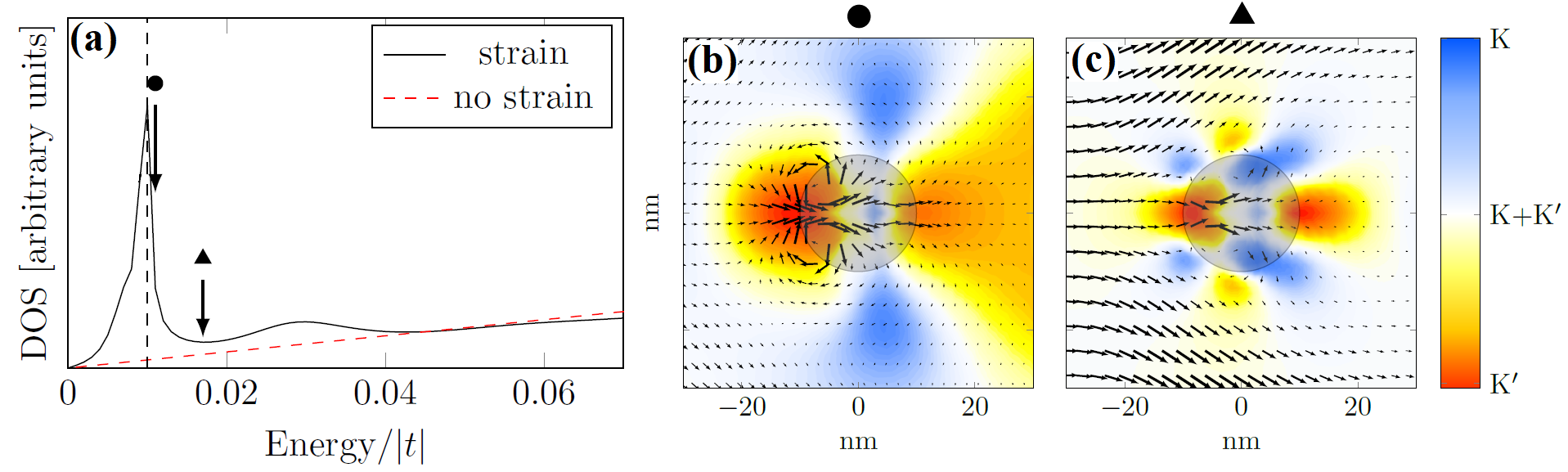}}
	\vspace*{8pt}
	\caption{(a) Average density of states in the Gaussian bump. The resonance energy of the Gaussian bump is indicated by the vertical dashed line. Some of the off-resonance energies are marked by symbols (circle and triangle). (b, c) Valley filtering effects corresponding to the two off-resonance energies. Figure adapted from Ref.~\citen{Settnes} with permission. \label{Fig:Settnes_valley_filter_drawback}}
\end{figure}

Arguably the most detrimental problem of using bumps for valley filtering purposes is the strong sensitivity of the efficiency of valley polarization to the incident energy. As we have mentioned in Sect.~\ref{Sect:LDOS}, the presence of deformations will alter the charge distribution, thus modulations of the density of states occur. Fig.~\ref{Fig:Settnes_valley_filter_drawback}(a) presents the average density of states in the presence (black) and absence (red) of a Gaussian bump, where one can identify a narrow and sharp resonance peak near $E=0.01t$ when the bump exists. The strong valley polarization effect shown in Fig.~\ref{Fig:Settnes_valley_filter}(d) is achieved right at the resonance energy. Even just by tuning the energy slightly as indicated by the dot in Fig.~\ref{Fig:Settnes_valley_filter_drawback}(a), the efficiency of the valley polarization effect is significantly reduced (Fig.~\ref{Fig:Settnes_valley_filter_drawback}(b)). The polarization becomes more localized and exhibits complicated angular dependence as the energy moves further away from the resonance (Fig.~\ref{Fig:Settnes_valley_filter_drawback}(c)). 
Furthermore, as the resonance energy is closely related to the geometric parameters of the bump, the resonance energy and thus ideal polarization is challenging to achieve without precise knowledge of the shape of the bump.
In the scattering formalism, the partial polarizations ($P<1$) and rather localized angular distribution observed in the majority of the scattering events are consistent with the tight-binding results for off-resonant energies, but one can not explore the resonance effect in a simple manner with the Born approximation. The problem requires a full treatment of the scattering matrix and becomes highly dependent on the geometry of the deformation.
\begin{figure}[ht]
	\centerline{\includegraphics[width=3.5in]{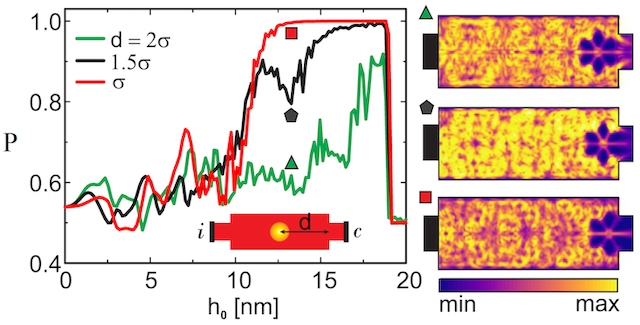}}
	\vspace*{8pt}
	\caption{Valley polarization $P$ for the two-terminal system shown in the inset versus the height of the bump for three different positions of the bump with respect to the right contact. The right panel shows current intensity plots for selected points indicated in the left figure. Figure adapted from Ref.~\citen{PeetersBump} with permission.\label{Fig:Peeters_valley_filter}}
\end{figure}

Another drawback of the bump lies in its localized structure that allows unperturbed currents around the deformed area. As a consequence, the composition of the transmitted current strongly depends on the location of the bubble with respect to the contact where current is collected.\cite{PeetersBump} Fig.~\ref{Fig:Peeters_valley_filter} presents results of the valley polarization as function of the location and height of the bump. The polarization is found to remain moderate as the distance between the bubble and the contact is too large. The requirement of a close proximity of the bubble to the contact inevitably imposes severe technical difficulties in sample fabrication.

\subsection{Fold geometry}\label{Sect:Valley_filter_fold}
In this section we discuss in detail the valley filtering properties of an extended deformation, i.e. a Gaussian fold (Fig.~\ref{Fig:deformations}(b)).\cite{MyPaperValleyFilter} The characteristics of the filter are analyzed as the incident angle and energy of the current are changed. In the discussion below, we present results in the absence of the strain-induced scalar potential, whose main effect is breaking the particle-hole symmetry (Readers can refer to Refs.~\citen{MyPaperValleyFilter,ZhaiDissertation} for more discussions on the effect of the scalar potential.).

As will be shown, among these two geometries in Fig.~\ref{Fig:deformations}, folds (or equivalent extended non-uniform strain geometries) are better valley filter devices in terms of degree of polarization and sizable transmitted currents. The reason for an improved performance relies on the extended geometry (incident waves can not move around it), and the robust resonances imposed by the structure that contributes to an optimal spatial separation between valley currents.

\subsubsection{Transverse transport across the fold}

\begin{figure}[ht]
	\centerline{\includegraphics[width=3.5in]{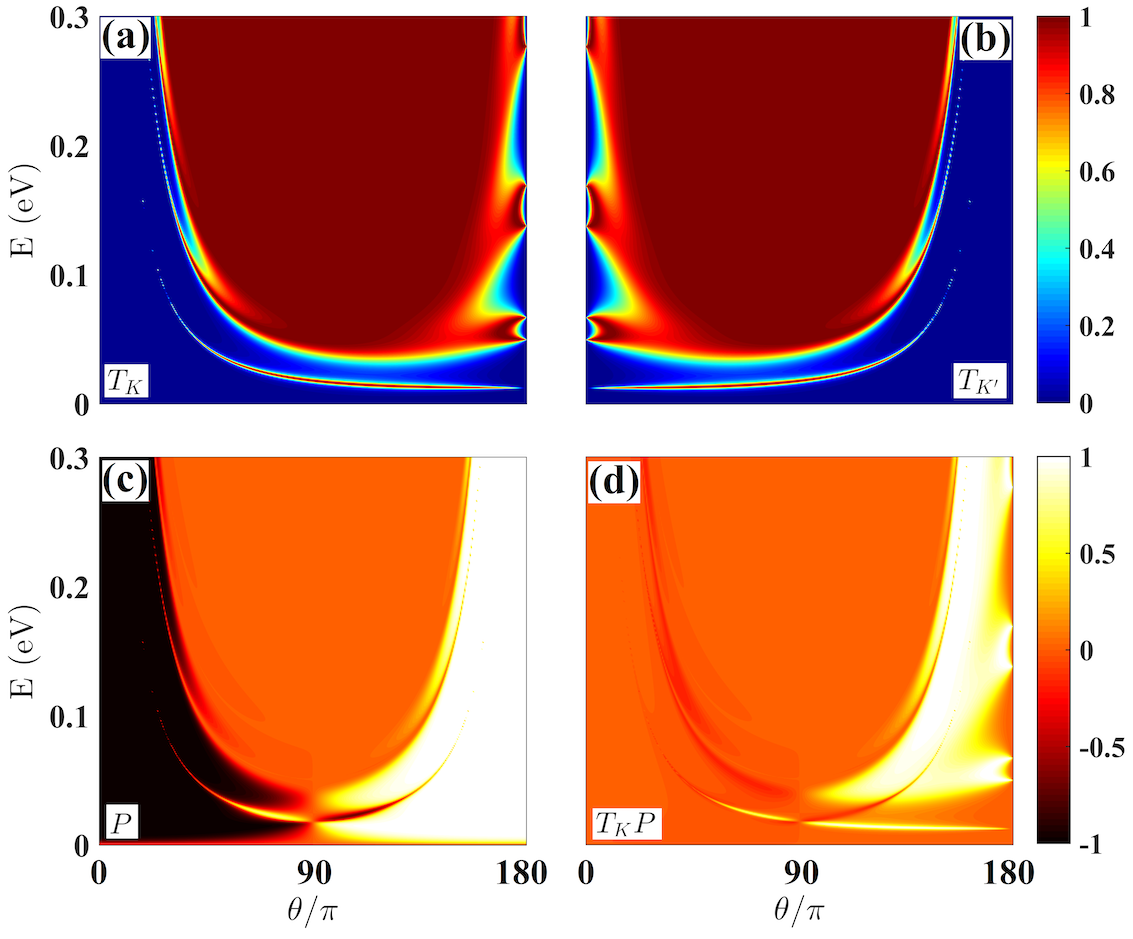}}
	\vspace*{8pt}
	\caption{(a, b) Transmission probability versus energy $E$ and incident angle $\theta$ for $K$ (left) and $K'$ (right)  valleys. (c, d) Valley polarization $P$ (left) and $T_KP$ (right) versus incident angle $\theta$ and energy $E$. Other parameters: $b=50$ nm, $\eta=0.1$. Figure adapted from Ref.~\citen{MyPaperValleyFilter} with permission. \label{Fig:Transmission_K_vs_Kprime}}
\end{figure}
The first row of Fig.~\ref{Fig:Transmission_K_vs_Kprime} shows the transmission probability of crossing the fold $T_K$ and $T_{K'}$ for electrons from valleys $K$ and $K'$ respectively, as functions of the incident angle $\theta$ (measured with respect to the fold axis) and energy.
Clearly, valley polarization is observed, i.e. $T_K(E,\theta)\ne T_{K'}(E,\theta)$, for a wide range of $\theta$ and energy. As the pseudo-vector potential respects time-reversal symmetry, transmission probabilities satisfy $T_K(E,\theta)=T_{K'}(E,180^\circ-\theta)$. In addition to the solid color regions, sharp lines in panels (a) and (b) with finite transmission are distinguishable. These correspond to energies in the `tunneling regime', i.e. states with imaginary wave vectors in the fold region. Apart from the robustness of the valley polarization effect against changes in energy and incident angle, it has also been shown that the effect remains pronounced even if strain strength (characterized by $\eta=h_0/b$) and fold width $b$ are varied\cite{MyPaperValleyFilter}.

\begin{figure}[ht]
	\centerline{\includegraphics[width=4in]{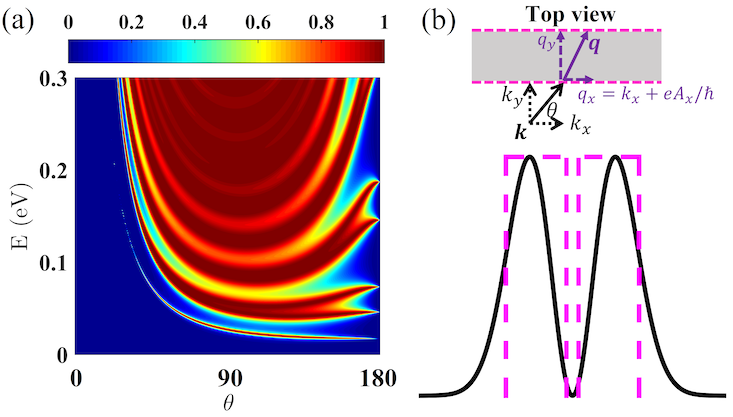}}
	\vspace*{8pt}
	\caption{(a) Transmission probability versus incident angle $\theta$ and energy $E$ for $K$ valley for a double square barrier potential. (b) The bottom plot shows the profile of $\bA(y)$ (solid black) and a double square potential (dashed magenta). The top plot shows schematics of the wave vectors inside and outside the potential barrier. Parameters: $b=50$ nm, $\eta=0.1$, width of the square barrier $=b$, separation $=0.2b$. Figure adapted from Ref.~\citen{MyPaperValleyFilter} with permission. \label{Fig:Double_barrier}}
\end{figure}

The transmission spectra can be understood through a double pseudo-vector potential barrier model. Fig.~\ref{Fig:Double_barrier}(a) shows the transmission spectrum for a double barrier potential, where panel (b) shows the barrier potential (dashed magenta) superimposed to the real pseudo-vector potential profile (solid black) and schematics of the wave vectors inside and outside the potential. The wave functions in different regions share the same form as Eq.~(\ref{Eq:Psi_extended}),
with $q_x$ and $q_y$ replaced by $k_x=|k|\cos\theta$ and $k_y=|k|\sin\theta$ outside the potential. By assuming that the electrons are incident from one side of the potential and employing the scattering matrix method, one can obtain the transmission probability easily.\cite{MyPaperValleyFilter} As the pseudo-vector potential exhibits opposite signs in the two valleys, the wave vectors of electrons from the two valleys are different in the potential (Fig.~\ref{Fig:Double_barrier}(b)), thus $T_K(E,\theta)\ne T_{K'}(E,\theta)$ in general.
The transmission spectrum of the double pseudo-vector potential barrier appears oscillatory along the energy axis due to the sharp edges, however it reproduces the main features shown in Fig.~\ref{Fig:Transmission_K_vs_Kprime}(a), especially resonances near $\theta=180^\circ$ (similar results are also presented in Fig.~\ref{Fig:Double_A_barrier_transmission_and_bound_states}). It is obvious, especially at low energy, that the resonances appear in pairs. The locations of the lower branch of the resonance pairs can be evaluated from the resonance condition of a single barrier structure $q_yb=n\pi$, where $q_y$ is the $y$ component of the wave vector in the barrier and $b$ is the width of the barrier. Each of these resonances splits and form a pair due to the coupling of the two barriers in the double-barrier structure (compare with Fig.~\ref{Fig:Single_A_barrier_transmission_and_bound_states}).

To characterize the efficiency of the Gaussian fold for inducing valley polarized transmitted currents, one can define the angle dependent polarization coefficient in terms of transmission coefficients as:
\begin{equation}
P=\frac{T_K-T_{K'}}{T_K+T_{K'}}
\end{equation}
Fig.~\ref{Fig:Transmission_K_vs_Kprime}(c) presents results for  $P$ corresponding to data shown in Fig.~\ref{Fig:Transmission_K_vs_Kprime}(a, b). $P$ exhibits a mirror symmetric structure with respect to $\theta=90^\circ$ due to time-reversal symmetry.
Large polarization regions appear at large and small incident angles with respect to the fold axis. 

Analogous to the case of the bump deformation, one may notice that the degree of polarization is not enough to ensure a measurable polarized current. In addition to high polarization values, a detectable signal must have a large transmission probability. Thus, it is convenient to evaluate the product of these two quantities as an indicator of the efficacy of the fold as a valley polarizer. Fig.~\ref{Fig:Transmission_K_vs_Kprime}(d) shows the product of the transmission probability $T$ and valley polarization $P$ for $K$ valley.
A large parameter region (golden area in Fig.~\ref{Fig:Transmission_K_vs_Kprime}(d)) with large values of both $T$ and $P$ can be identified, indicating high efficiency for a wide range of energies. $TP$ results for valley $K'$ can be obtained by applying the relation $T_{K'}P(180^\circ-\theta)=-T_{K}P(\theta)$, which results in an antisymmetric profile with respect to $\theta=90^\circ$.

\subsubsection{Drawbacks of extended folds for valley filtering}

\begin{figure}[ht]
	\centerline{\includegraphics[width=3.5in]{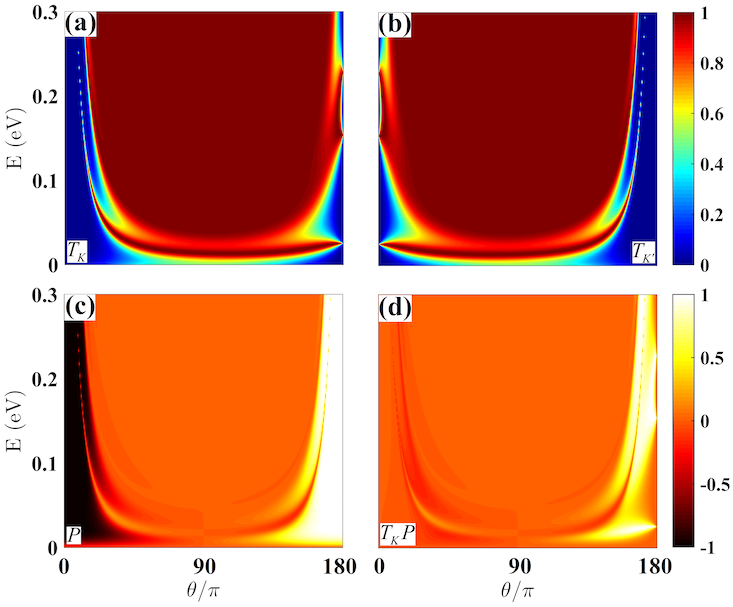}}
	\vspace*{8pt}
	\caption{Results for a Gaussian fold with axis oriented at an angle $\gamma=25^\circ$ with respect to the zigzag direction. (a, b) Transmission $T$ for $K$ and $K'$ valley, (c) Polarization $P$, and (d) $TP$ for $K$ valley. Other parameters: $b=50$ nm, $\eta=0.1$. Figure adapted from Ref.~\citen{MyPaperValleyFilter} with permission. \label{Fold_along_25_deg}}
\end{figure}

The main drawback of using extended folds for valley filtering is that the polarization and magnitude of transmitted current depends on the fold axis orientation. The influence of the orientation can be seen by considering an arbitrary direction $\gamma$ for the fold axis with respect to the zigzag direction. By choosing the $\hat{x}$ direction along the fold axis, the pseudo-magnetic field and the pseudo-vector potential in this case are obtained by applying the appropriate rotation, transforming as $\textbf{B}\rightarrow \textbf{B}\cos(3\gamma)$ and $A_x\rightarrow A_x\cos(3\gamma)$~\cite{RotationEffectStrainPRB2015,RotationEffectStrainPRB2016,MyPaperValleyFilter,RamonFold}. 

Fig.~\ref{Fold_along_25_deg} shows the results for a fold along $\gamma=25^\circ$. The large transmission region increases (red area in Fig.~\ref{Fold_along_25_deg}(a,b)) due to a smaller pseudo-vector potential barrier ($A_x\rightarrow A_x\cos(3\gamma)\approx0.26A_x$), and as a consequence, the extent of the valley polarized regime decreases (Fig.~\ref{Fold_along_25_deg}(c)). It is clear from the nature of the rotation that folds along the zigzag direction, e.g. $\gamma=0^\circ$, will yield the largest valley polarization effect. However, folds along the armchair direction, e.g. $\gamma=30^\circ$, can not produce valley polarized current at all.

\section{Future directions}\label{Sect:Future}
Although the valley Hall effect in bilayer graphene where inversion symmetry is broken has already been observed,\cite{ValleytronicsNatPhysBilayerGraphene,ValleytronicsNatPhysBilayerGraphene2} the valley filtering effect in monolayer graphene remains unreported. With the rapid advance of experimental techniques for fabrication of high quality samples, one may achieve valley filtering in graphene with multiple folds.\cite{MyPaperValleyFilter} For instance, one can employ a double fold structure, where the first fold acts as a polarizer, while the second acts as a detector. By choosing the direction of the incident current properly, the intensity of the current -on average- is expected to decrease by half after passing through the first fold due to the valley filtering effect, in analogy with optical polarizers. The transmitted current is fully valley polarized and the second fold will allow full transmission in the parallel configuration (It can also help eliminate any remaining component of the opposite valley if the transmitted current after the first fold is partially polarized.).  By adjusting the direction of the axis of the second fold, partial or null transmission can be achieved too.

Interesting transport phenomena can also be explored in terms of the interplay of external magnetic field and strain-induced pseudo-magnetic field.\cite{SettnesRealandPseudoMagnetic}
As the external magnetic field affects the two valleys equally, while the pseudo-magnetic field exhibits opposite signs, the balance between the two valleys will be broken. An ideal proposal for valley filtering considers adding an external magnetic field with the same magnitude as the constant pseudo-magnetic field in triaxially strained graphene, where electrons in one valley experiences vanishing magnetic field, while those in the other valley feel a magnetic field twice the intensity of the external field.\cite{SettnesRealandPseudoMagnetic} Moreover, in the presence of a strong external magnetic field, the bulk bands of the system will be gapped by the development of Landau levels. One can then explore the transport phenomena due to the confined states that propagate only along the folds discussed in Sect.~\ref{Sect:Confinement} by positioning the chemical potential in one of the Landau level gaps.\cite{RamonFold}

It is also interesting to combine strain with other tunable factors, e.g. Moir\'e patterns in bilayer structures\cite{LiangFuStrainMoire,TongQingjunNatPhys2017}, to control the property and functionality of the materials. Recently, it has been proposed that the effect of interlayer coupling on electrons in twisted bilayer graphene can be understood in terms of pseudo-magnetic fields.\cite{PseudoBFieldMoirePRB2019,PseudoBFieldMoirePRL2012}  Interesting effects in the presence of pseudo-magnetic fields originated from both strain and Moir\'e patterns can be expected. Recent studies in twisted bilayer graphene report that strain can enlarge the separation of the flat conduction and valence bands as well as set a lower bound for the Fermi velocity.\cite{LiangFuStrainMoire}

Usually electron-electron interactions in graphene~\cite{eeInteractionRMP2012} can be neglected and a single particle picture is enough to understand most of the physical phenomena observed. However, they are found to be important under certain circumstances, among which the fractional quantum Hall effect in the presence of strong magnetic fields is an example~\cite{FractionalQHE1,FractionalQHE2}. In Sect.~\ref{Sect:LDOS} we have shown that strain can help manipulate charge density distribution either on the sublattice level or globally in the deformed area. Especially, in Sect.~\ref{Sect:Confinement} we have seen that flat bands can occur in graphene folds at low energy. As the kinetic energy of electrons in such flat energy levels is dramatically reduced, electron-electron interactions might become dominate. Therefore, strained graphene may function as a relatively simple setup for exploring electron correlation physics\cite{SuperconductivityTwistedBilayer,CaoYuanCorrelated}.

\section*{Acknowledgments}
We acknowledge support from NSF Grant No.\ DMR-1508325.

\bibliographystyle{ws-mplb}
\bibliography{ReviewRefs}

\end{document}